\documentclass[aps,prd,amsmath,amssymb,superscriptaddress,twocolumn,nofootinbib,showpacs,preprintnumbers,notitlepage]{revtex4-1}
\usepackage[usenames,dvipsnames,svgnames,table]{xcolor}
\usepackage{graphicx}
\usepackage{dcolumn}
\usepackage{bm}
\usepackage{hyperref}
\hypersetup{ 
    pdfnewwindow=true,      
    colorlinks=true,       
    allcolors=[RGB]{31 119 180}
} 
\usepackage[capitalise]{cleveref}
\usepackage{dsfont}
\usepackage{placeins}
\usepackage{todonotes}
\usepackage[utf8]{inputenc}
\usepackage{gensymb}
\usepackage{multirow}
\usepackage{amsmath}
\usepackage{amsfonts}
\usepackage{amssymb}
\usepackage{enumitem,amssymb}
\usepackage{array}   
\newcolumntype{L}{>{$}l<{$}} 
\newcolumntype{R}{>{$}r<{$}}
\newcolumntype{C}{>{$}c<{$}}
\usepackage{xspace}
\usepackage{braket}
\usepackage{units}
\usepackage{slashed}
\usepackage[normalem]{ulem}
\usepackage[format=plain,justification=RaggedRight,singlelinecheck=false]{caption}
\usepackage[format=plain,justification=centering,singlelinecheck=false]{subcaption}
\usepackage{tabularx}
\usepackage{nameref}

\usepackage{xpatch}
\makeatletter
\xpatchcmd{\@ssect@ltx}{\@xsect}{\protected@edef\@currentlabelname{#8}\@xsect}{}{}
\xpatchcmd{\@sect@ltx}{\@xsect}{\protected@edef\@currentlabelname{#8}\@xsect}{}{}
\makeatother
\usepackage{hyperref}

\graphicspath{{figures/}}

\let\Re\relax

\DeclareMathOperator{\Re}{Re}

\newcommand{\eg}{{\it e.g.}\xspace}
\newcommand{\cf}{{\it cf.}\xspace}

\newcommand{\ie}{{\it i.e.}\xspace}

\newcommand{\mevnospace}{\ensuremath{{\mathrm{\,Me\kern -0.1em V}}}}
\newcommand{\gevnospace}{\ensuremath{{\mathrm{\,Ge\kern -0.1em V}}}}
\newcommand{\tevnospace}{\ensuremath{{\mathrm{\,Te\kern -0.1em V}}}}

\newcommand{\gev}{\gevnospace\xspace}
\newcommand{\gevsq}{\ensuremath{\gevnospace^2}}

\newcommand{\fm}{\ensuremath{{\mathrm{\,fm}}}}

\newcommand\bsub{\begin{subequations}}
\newcommand\esub{\end{subequations}}

\newcommand{\helE}{\ensuremath{{\sigma_J}}}

\newlist{todolist}{itemize}{2}
\setlist[todolist]{label=$\square$}
\usepackage{pifont}

\definecolor{jpac-blue}{RGB}{31,119,180}
\definecolor{jpac-red}{RGB}{214,39,40}
\definecolor{jpac-green}{RGB}{0,158,115}
\definecolor{jpac-orange}{RGB}{255,127,14}

\usepackage{orcidlink}

\usepackage{tikz}
\usepackage[compat=1.1.0]{tikz-feynman}

\tikzset{
    feynman scale/.style={
        transform shape,
        scale={#1},
        every node/.append style={
            scale={1/#1}
        },
    }
}

\usepackage{cancel}

\newcommand{\AGH}{AGH University of Krakow, Faculty of Physics and Applied Computer Science, PL-30-059 Krak\'ow, Poland}
\newcommand{\catania}{INFN Sezione di Catania, I-95123 Catania, Italy}
\newcommand{\ceem}{Center for  Exploration  of  Energy  and  Matter, Indiana  University, Bloomington,  IN  47403,  USA}
\newcommand{\indiana}{Department of Physics, Indiana  University, Bloomington,  IN  47405,  USA}
\newcommand{\jlab}{Theory Center, Thomas  Jefferson  National  Accelerator  Facility, Newport  News,  VA  23606,  USA}
\newcommand{\messina}{Dipartimento di Scienze Matematiche e Informatiche, Scienze Fisiche e Scienze della Terra, Universit\`a degli Studi di Messina, I-98122 Messina, Italy}
\newcommand{\HISKP}{Helmholtz-Institut f\"{u}r Strahlen- und Kernphysik (Theorie) and Bethe Center for Theoretical Physics, Universit\"{a}t Bonn, D-53115 Bonn, Germany}
\newcommand{\ub}{Departament de F\'isica Qu\`antica i Astrof\'isica and Institut de Ci\`encies del Cosmos, Universitat de Barcelona, E-08028 Barcelona, Spain}
\newcommand{\uned}{Departamento de F\'isica Interdisciplinar, Universidad Nacional de Educaci\'on a Distancia (UNED), E-28040 Madrid, Spain}
\newcommand{\gwu}{Department of Physics, The George Washington University, Washington, DC 20052, USA}
\newcommand{\ucb}{Department of Physics, University of California, Berkeley, CA 94720, USA}
\newcommand{\lbnl}{Nuclear Science Division, Lawrence Berkeley National Laboratory, Berkeley, CA 94720, USA}
\newcommand{\odu}{Department of Physics, Old Dominion University, Norfolk, VA 23529, USA}

\begin{document}

\preprint{JLAB-THY-24-4129}
\title{Revisiting gauge invariance and Reggeization of pion exchange}

\author{Gloria~\surname{Monta\~na}\orcidlink{0000-0001-8093-6682}}
\email{gmontana@jlab.org}
\affiliation{\jlab}
\author{Daniel~\surname{Winney}\orcidlink{0000-0002-8076-243X}}
\email{winney@hiskp.uni-bonn.de}
\affiliation{\HISKP}
\author{{\L}ukasz Bibrzycki\orcidlink{0000-0002-6117-4894}}
\affiliation{\AGH}
\author{C\'esar~\surname{Fern\'andez-Ram\'irez}\orcidlink{0000-0001-8979-5660}}
\affiliation{\uned}
\author{Giorgio~\surname{Foti}\orcidlink{0009-0000-9791-3823}}
\affiliation{\messina}
\affiliation{\catania}
\author{Nadine~\surname{Hammoud}\orcidlink{0000-0002-8395-0647}}
\affiliation{\ub}
\author{Vincent~Mathieu\orcidlink{0000-0003-4955-3311}}
\affiliation{\ub}
\author{Robert~J.~\surname{Perry}\orcidlink{0000-0002-2954-5050}}
\affiliation{\ub}
\author{Alessandro~\surname{Pilloni}\orcidlink{0000-0003-4257-0928}}
\affiliation{\messina}
\affiliation{\catania}
\author{Arkaitz~\surname{Rodas}\orcidlink{0000-0003-2702-5286}}
\affiliation{\jlab}
\affiliation{\odu}
\author{Vanamali~\surname{Shastry}\orcidlink{0000-0003-1296-8468}}
\affiliation{\indiana}
\affiliation{\ceem}
\author{Wyatt~A.~\surname{Smith}\orcidlink{0009-0001-3244-6889}}
\affiliation{\gwu}
\affiliation{\ucb}
\affiliation{\lbnl}
\author{Adam~P.~\surname{Szczepaniak}\orcidlink{0000-0002-4156-5492}}
\affiliation{\jlab}
\affiliation{\ceem}
\affiliation{\indiana}
\collaboration{Joint Physics Analysis Center}
\begin{abstract}
The Reggeized pion is expected to provide the main contribution to the forward cross section in light meson photoproduction reactions with charge exchange at high energies.
We discuss the Reggeization of pion exchange in charged pion photoproduction with an emphasis on consistency with current conservation. We show that the gauge-invariant amplitude for the exchange of a particle with generic even spin $J\geq 2$ in the $t$-channel is analytic at $J=0$, and that it can be interpreted in terms of the nucleon electric current. This enables us to reconcile the dynamics in the $s$- and $u$-channel, which involves also nucleon exchanges, with the amplitude expressed in terms of $t$-channel partial waves, as required by Regge theory.
\end{abstract}
\maketitle

\section{Introduction}
\label{sec:intro}

Understanding the idiosyncrasies of pion exchange in hadronic reactions is of fundamental interest and has been widely studied. 
Low-energy pion interactions with hadronic matter are constrained by chiral symmetry which allows the derivation of the effective interactions, for example to describe nucleon-nucleon potentials~\cite{Machleidt:1987hj,Ericson:1988gk,Bernard:1992nc}. 
At high center-of-mass energy and small momentum transfer, hadronic scattering amplitudes are known to be described by the exchange of Reggeons in the $t$-channel~\cite{Regge:1959mz,Collins:1977jy,Irving:1977ea}. 
Generally, natural-parity trajectories give the largest contribution to the cross section. However, for charge-exchange reactions, the pion trajectory dominates the very forward region. 
More recently, it has become important to understand pion exchange 
due to its possible role in the phenomenology of the $XY\!Z$ states observed in the charmonium spectrum~\cite{Albaladejo:2020tzt,Winney:2022tky} and as a benchmark in the photoproduction of light exotic hybrid mesons~\cite{Close:1994pr,Afanasev:1999rb,Szczepaniak:2001qz}.

So far, pion exchange has been studied with phenomenological approaches. 
Given new and forthcoming high-precision data from JLab for a variety of photo- and electroproduction reactions in which it 
plays a pivotal role, it is essential to revisit the pion exchange mechanism in a more fundamental approach.   

In this paper, we consider the photoproduction of a single charged pion,
\begin{equation}\label{eq:s-channel-reaction}
 \gamma(k) + N(p_i) \to \pi^\pm(p_\pi) + N'(p_f) \ ,
\end{equation}
where the four-momenta are indicated in parenthesis, and $m_N$ and $m_\pi$ the mass of the nucleon and the pion, respectively.\footnote{Throughout the paper, the isospin-breaking mass difference between the proton and the neutron is neglected.} 
This reaction offers the cleanest probe of pion exchange due to the point-like nature of the photon interaction. The kinematics of this reaction are summarized in \cref{app:Born-s}. The consequences of gauge invariance in photon-hadron interactions are often non-trivial~\cite{Borasoy:2005zg,Borasoy:2007ku}.
The issue of gauge invariance in the presence of a Reggeized pion has been extensively discussed in the past (see \eg Refs.~\cite{PhysRevLett.21.250,Kellett:1970fdw,Jones:1979aa,Rahnama:1990mt,Guidal:1997hy,Sibirtsev:2007wk,Sibirtsev:2009bj,Yu:2011zu}). Typically, a gauge-invariant Born-type model is constructed and a recipe to replace the pion propagator by a Regge propagator is employed. This Reggeization procedure, however, has never been studied rigorously.

The way we will approach this problem is by expanding in terms of partial waves for the crossed $t$-channel reaction, 
\begin{equation}\label{eq:t-channel-reaction}
 \gamma(k) + \pi^\mp(-p_{\pi}) \to  \bar N(-p_{i}) + N'(p_f) \ ,
\end{equation}
where the allowed $t$-channel exchanges can be readily identified. \cref{app:Born-t} summarizes the kinematics of this reaction.
The partial-wave expansion of the amplitude is most easily written in the helicity formalism and reads
\begin{equation}\label{eq:PWexp-t}
    A_{\lambda_\gamma\lambda_i\lambda_f}(s,t)= \sum_{J}(2J+1) \, a^J_{\lambda_\gamma\lambda_i\lambda_f}(t) \, d^J_{\lambda_\gamma\lambda_i-\lambda_f}(\theta_t) \ ,
\end{equation}
where $\lambda_\gamma$ and $\lambda_{i,f}$ are the $t$-channel helicities of the photon and the nucleons, and $d^J_{\lambda_\gamma,\lambda_i-\lambda_f}(\theta_t)$ are the Wigner rotation matrices, as a function of the $t$-channel scattering angle $\theta_t$. The functions $a^J_{\lambda_\gamma\lambda_i\lambda_f}(t)$ are the partial waves. Although the scattering amplitude should obey crossing symmetry, as written, the partial wave decomposition cannot be directly continued from the physical region of \cref{eq:t-channel-reaction} to that of \cref{eq:s-channel-reaction}. Instead, the $t$-channel partial-wave series must first be re-summed and then analytically continued to the $s$-channel physical region.  
The resummation of infinitely many partial waves gives rise to the Reggeons expected in Regge theory~\cite{Gribov:2009cfk}. 
The parameterization of each partial wave as a simple particle exchange is sometimes referred to as the Feynman-Van~Hove picture, whereby Reggeons represent the exchange of an infinite tower of particles with increasing spin~\cite{vanHove:1967zz,Ravndal:1970xe}. The construction of the partial waves and their summation is the central goal of this work.

The transverse nature of the photon implies that the value of $J$ is constrained to be $J \ge |\lambda_\gamma| = 1$ in~\eqref{eq:PWexp-t}.
That means that there is no $J=0$ component that can accommodate pion exchange. Having an explicit pion contribution requires us to extend the sum to unphysical values of $J$. We thus write a model for an exchange of particles with definite spin $J$ that can be analytically continued to $J=0$.
The model is written as a product of two separate vertices, which are constructed from Lorentz-covariant tensor structures of the corresponding four-momenta and helicity operators. 

The number of independent Lorentz structures is finite and determined by the allowed quantum numbers in the photon-pion-Reggeon vertex. These independent structures allow gauge invariance to be enforced at the vertex level, which then guarantees that each partial wave and therefore the re-summed amplitude also satisfies gauge invariance. 
In the center-of-momentum (CM) frame of the $t$-channel reaction, one can see that the $J=0$ exchange is forbidden by helicity conservation. Making sense of a pion exchange requires some deeper understanding of the amplitude.

The main novelty of this paper is the following: When considered as an analytic function of $J$, the gauge-invariant partial waves contain a singularity at $J=0$.
This singularity is canceled by a zero of the Wigner function at $J=0$. We will show that the resulting amplitude contains the pion pole. 
This new contribution can be added to the infinite sum of partial waves, which gives a Reggeized amplitude for pion exchange that is consistent with gauge invariance. We find a $s$ dependence consistent with na\"ive Regge expectations, but a more complicated $t$ dependence than that obtained using the Reggeization prescription employed in previous studies~\cite{Kellett:1970fdw,Jones:1979aa,Rahnama:1990mt,Guidal:1997hy,Sibirtsev:2007wk,Yu:2011zu}.

In the Born-like models commonly employed at lower energies, the lack of gauge invariance of the pion-exchange diagram alone is not an issue: The amplitude for charged pion photoproduction arises from the exchange of both pions in the $t$-channel and nucleons in the $s$-channel\footnote{We use the notation ``$s$-channel'' and ``$t$-channel'' to denote the Born terms with poles in the corresponding Mandelstam invariants, but also to refer to the CM frame of the reaction where the Mandelstam invariant has the meaning of the CM energy of the colliding particles. These are common notations in the literature. We specify what is the appropriate meaning (Born term or reference frame) each time in the text to minimize confusion.} (or $u$-channel, depending on the charge of the pion), and the total amplitude is gauge invariant. In the second part of this paper, we will make a connection with these low-energy effective descriptions of pion photoproduction. We will see that the Born diagrams are related to each other and that the contribution of the pion pole is actually contained in the nucleon exchange term of the Born amplitude in the $t$ channel CM frame. By studying this Born-level model, we show that a na\"ive isolation of the pion exchange process produces an amplitude which is not gauge invariant, and additionally a cross section which is not Lorentz invariant. As a result, one obtains different answers when computing it in different reference frames. In particular, it vanishes in the CM frame of the $t$-channel, relevant to Reggeization, because a pion cannot decay into a pion and a photon if helicity is to be conserved. This demonstrates again the importance of working with gauge invariant amplitudes. 

The rest of the paper is organized as follows. In Section~\ref{sec:reggeization}, we describe in detail the process of constructing a Reggeized pion exchange amplitude consistent with gauge invariance, in Section~\ref{sec:born}, we discuss the effective Lagrangian approach to this problem and show that the nucleon Born term contains the pion pole, in Section~\ref{sec:results} we compare and contrast the different Regge prescriptions. Finally, in Section~\ref{sec:conclusions}, we summarize the results of this paper and suggest further work. 

\section{Reggeization of pion exchange}
\label{sec:reggeization}

According to Regge theory, the dynamics of peripheral photoproduction reactions are determined by the near-threshold singularities in the crossed ($t$-) channel.
As discussed, pion exchange is the dominant contribution at small momentum transfer, $t$.
Reggeizing the pion exchange involves accounting for all the higher spin particles that lie on the Regge trajectory of the pion, in addition to the pion itself. This can be achieved by summing the exchanges in the $t$-channel with spins \mbox{$J=0,2,4,\ldots$} and negative parity. 

This procedure has often not been followed. Instead, an empirical prescription has been employed, that we name VGL after the work of Vanderhaeghen, Guidal and Laget~\cite{Guidal:1997hy}. In this approach, a gauge-invariant Born-level amplitude is simply multiplied by an overall factor, 
\begin{align}\label{eq:prescrip}
&A^\text{VGL}_{\lambda_\gamma\lambda_i\lambda_f}(s,t)   \\
&\qquad =A^\text{Born}_{\lambda_\gamma\lambda_i\lambda_f}(s,t) \,(t-m_\pi^2) \,\alpha^\prime \, \Gamma(-\alpha(t)) \left(\frac{s}{s_0} \right)^{\alpha(t)} ~,\nonumber 
\end{align}
where $\alpha(t)=\alpha'(t-m_\pi^2)$ is a real, linear pion trajectory and $s_0 = 1$ GeV$^2$ is a characteristic energy scale. This is interpreted as ``Reggeizing the pion'' but there is no fundamental theoretical justification for this prescription. Its use should be understood as a phenomenological recipe.

Here, we take a fundamental approach to the study of Reggeized pion exchange. Specifically, we consider the tower of exchanges explicitly by first constructing the $t$-channel partial waves consistent with current conservation, and then summing the partial-wave series. The large-$s$ Regge amplitude is finally obtained by analytic continuation to the $s$-channel physical region.

\subsection{Exchange of arbitrary spin ($J\geq 2)$}\label{subsec:a}

In general, a $t$-channel exchange can have any of four distinct spin-parity combinations: $J^{P}=(\textrm{even})^-$, $(\textrm{odd})^-$, $(\textrm{even})^+$, and $(\textrm{odd})^+$. This, in principle, would allow one to study trajectories with different signature $(-1)^J$ and naturality $P(-1)^J$, separately. Because we are interested in the pion, we consider only unnatural exchanges with even signature. An analogous approach can be used to describe the other exchanges.

\begin{figure}[h]
\begin{subfigure}[b]{0.4\columnwidth}
\centering
\begin{tikzpicture}[feynman scale=0.8, baseline=(l1.base)]
    \begin{feynman}[small]
      \vertex (v1) at (0,0);
      \vertex (v2) at (1,0);
      \vertex (l1) at (-0.9,0.8) {\( \gamma \)};
      \vertex (l2) at (-0.8,-0.8) {\( \bar \pi \)};
      \diagram*{
        (l1) -- [photon, thick] (v1), 
        (l2) -- [scalar, thick] (v1),
        (v1) -- [jpac-blue,ultra thick, photon, edge label=\( \textcolor{black}{J^{P}}\ \)] (v2),
       };
       \draw[blob, jpac-blue] (v1) circle(0.08) ;
       \end{feynman}
       \draw [jpac-red,->] (-1.9,0) -- (-1.5,0) node [pos=0.5,above,font=\footnotesize] {\(t\)};
  \end{tikzpicture}
\end{subfigure}
\begin{subfigure}[b]{0.5\columnwidth}
\centering
  \begin{tikzpicture}[feynman scale=0.8,baseline=(l1.base)]
    \begin{feynman}[small]
      \vertex (v1) at (-1,0);
      \vertex (v2) at (0,0);
      \vertex (r1) at (0.8,0.8) {\( \bar N \)};
      \vertex (r2) at (0.8,-0.8) {\( N \)};
      \diagram*{
        (v2) -- [plain, thick] (r1), 
        (v2) -- [plain, thick] (r2),
        (v1) -- [jpac-blue,ultra thick,photon ,edge label=\( \textcolor{black}{J^{P}}\ \)] (v2),
       };
       \draw[blob, jpac-blue] (v2) circle(0.08) ;
       \end{feynman}
       \draw [jpac-red,->] (-1.9,0) -- (-1.5,0) node [pos=0.5,above,font=\footnotesize] {\(t\)};
  \end{tikzpicture}
\end{subfigure}
\caption{Vertices of the pion photoproduction reaction with the exchange of a particle with $J^P$ in the $t$-channel. }
\label{fig:vertices}
\end{figure}
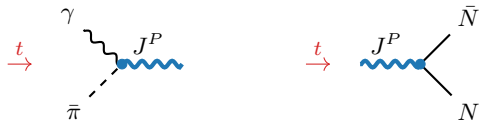

Based on the idea of Regge factorization of the $t$-channel residues, we split the $t$-channel helicity partial waves into the two vertices as shown in \cref{fig:vertices}. The vertices are constructed in terms of Lorentz covariant objects and can only depend on the three independent four-momenta: $k^\mu$, $p_\pi^\mu$, $P^\mu=p_i^\mu+p_f^\mu$, as well as the polarization tensors of the photon, $\epsilon_\mu(k,\lambda_\gamma)$, and of the exchange, $\epsilon_{\mu_1\cdots\mu_J}(k-p_\pi, \helE)$ (see \eg~\cite{Ravndal:1970xe}). For brevity, we omit the dependence on exchange momentum $k-p_\pi$. We consider the reference frame in which the exchanged particle is at rest (\ie the $t$-channel CM frame). In this frame, $\helE$ is the spin projection along the $z$-axis, which is chosen along the direction of the photon. 

The $\gamma\pi$ system has a minimal (\ie $S$-wave) spin-parity of $1^+$ (\ie $1^-\otimes 0^-=1^+$). 
Thus $\gamma\pi$ can couple to particles on unnatural trajectories with $J\geq1$ with two different Lorentz structures corresponding to orbital $L = J\pm 1$. Current conservation can be imposed at the vertex level, so that gauge invariance is automatically satisfied.

The two independent structures are 
\begin{align}
\label{eq:vertex-gammapi-unnat-traj}
  V^J_{\lambda_\gamma}(&\helE)  =  2\sqrt{2} \, g_{\gamma\pi} \, \epsilon^*_{\nu_1\cdots\nu_J}(\helE) \, \epsilon_\mu(k,\lambda_\gamma) \\ 
&\times \left( k^{\nu_1} \cdots k^{\nu_{J-1}} \right) \, \Big[ k^{\nu_J} \, p_\pi^\mu -  g^{\nu_J\mu}\left(k \cdot p_{ \pi}\right)  \Big]  \nonumber  
\end{align} 
and 
\begin{align}
  \tilde{V}^J_{\lambda_\gamma}(&\helE)  =  2\sqrt{2} \, \tilde{g}_{\gamma\pi} \, \epsilon^*_{\nu_1\cdots\nu_J}(\helE) \, \epsilon_\mu(k,\lambda_\gamma) \\ 
&\times \left( k^{\nu_1} \cdots k^{\nu_{J}} \right) \, \Big[ \left(k \cdot p_{ \pi}\right) \, k^\mu -   k^2 \, p_\pi^\mu \Big]  \nonumber  \ .
\end{align} 
The latter vertex contributes only to virtual longitudinal photons and will not be discussed 
further. 
The coupling constant $g_{\gamma\pi}$ is proportional to the electric charge of the pion, $e_\pi=\pm e$, and has
mass dimensions $[g_{\gamma\pi}]=-(J+1)$. 

In \cref{eq:vertex-gammapi-unnat-traj}, one of the Lorentz indices of the exchange has to be contracted with the structure in square brackets. Thus this structure does not allow a $J=0$ exchange carrying no Lorentz indices.
Further, in the $t$-channel kinematics, the pion moves in the opposite direction to the photon and any other vertex would vanish for real photons, as $\epsilon(k,\lambda_\gamma)\cdot p_\pi=0$. Therefore, \cref{eq:vertex-gammapi-unnat-traj} is only defined for $J\geq 1$ and we will come back to the $J=0$ case later.

For the nucleon vertex, we write 
\begin{align} \label{eq:vertex-NNbar-unnat-traj-NF} 
    V^J_{\lambda_i\lambda_f}(\helE) = g_{N\bar N} \left(P^{\nu_1} \cdots P^{\nu_{J}}\right) \epsilon_{\nu_1\cdots\nu_J }(\helE) ~~& \\
    \times \, \bar u(p_f,\lambda_f) \,\gamma_5 \,v(-p_{i}, \lambda_i&) \ , \nonumber
\end{align}
which produces a $N\bar N$ pair with minimal spin-parity of $0^-$ and which carry same helicity. It is worth noticing that, when crossed to the $s$-channel, this vertex gives a spin-flip amplitude.
It is possible to consider another opposite-helicity (spin-nonflip in the $s$-channel) amplitude, on which we will comment later. 
From \cref{eq:vertex-NNbar-unnat-traj-NF}, the coupling $g_{N\bar N}$ must depend on $J$ and has mass dimensions $[g_{N\bar N}] = {-(J+1)}$. 
The Dirac spinors $u(p_f,\lambda_f)$ and $v(-p_{i},\lambda_i)$ and their normalization are given in \cref{eq:spinors-t}. 

In the Regge pole approximation, the amplitude describing the exchange of a spin-$J$ particle on the pion trajectory, $\alpha(t)$, is obtained by introducing a pole in angular momentum, $\left[J-\alpha(t)\right]^{-1}$. We define 
\begin{align}\label{eq:vertices_with_reggepole}
    A^J_{\lambda_\gamma\lambda_i\lambda_f}(s,t)&=\sum_{\helE} \frac{V^J_{\lambda_\gamma}(\helE)\, V^J_{\lambda_i\lambda_f}(\helE)}{J-\alpha(t)}
       \ ,
\end{align}
where the $s,t$ dependence is implicit in the vertices.
This construction does not rely on a functional form of $\alpha(t)$ which is assumed only to be analytic at $t=m_\pi^2$, and satisfying $\Re\alpha(m_\pi^2) = 0$. 
By isolating the angular dependence, it is possible to show that 
\begin{equation}
    \label{eq:AJ}
A^J_{\lambda_\gamma\lambda_i\lambda_f}(s,t)=a^J_{\lambda_\gamma\lambda_i\lambda_f}(t) \, d^J_{\lambda_\gamma\lambda_i-\lambda_f}(\theta_t)
\end{equation}
where, 
\begin{align} \label{eq:pwJ}
a^J_{\lambda_\gamma\lambda_i\lambda_f}(t) \equiv \frac{2 \, e_{\pi} \,g_J \,  c_J^2 \, t}{J-\alpha(t)}  (2\lambda_i\delta_{\lambda_i\lambda_f})\, \sqrt{\frac{J+1}{J}}(-2p_tk_t)^J  \ ,
\end{align} 
and $e_\pi \, g_J\equiv g_{\gamma\pi} \, g_{N\bar N}$.
The coefficient $c_J=J! \sqrt{2^J/(2J)!}$ arises from the contraction of $J$ momenta with the polarization tensor of the spin-$J$ particle (details are given in \cref{app:exchange}). Note the factor of $-t$ that makes the amplitudes vanish in the forward direction, as expected by helicity conservation. 

We can make use of the relationship between the Wigner $d$-functions and the Jacobi polynomials (summarized in \cref{app:sum}). In particular, because the two nucleons always have the same helicity in \cref{eq:vertex-NNbar-unnat-traj-NF}, \ie, $\lambda_i - \lambda_f = 0$, we may write:
\begin{equation}\label{eq:dJ}
    d^J_{\lambda_\gamma,0}(\theta_t) =
    \sqrt{\frac{J+1}{2J}} \, d^1_{\lambda_\gamma,0}(\theta_t) \, P^{11}_{J-1}(z_t) ~.
\end{equation}
Finally, we want to impose definite signature, \ie the fact that odd and even values of $J$ belong to independent Regge amplitudes. The Jacobi polynomials have the symmetry property, $P_n^{ab}(-z)=(-1)^nP_n^{ba}(z)$.
Thus one can replace $P_{J-1}^{11}(z_t) \to \frac{1}{2}\left[P_{J-1}^{11}(z_t) - P_{J-1}^{11}(-z_t)\right]$ to make the amplitude vanish at odd integer $J$. 

\subsection{Analytic continuation to $J=0$}
The amplitude in \cref{eq:vertices_with_reggepole} has been constructed by considering $J\geq 1$. Clearly \cref{eq:pwJ} breaks down at $J=0$ due to the kinematic pole at this point. However, for the sum in \cref{eq:PWexp-t} to represent the Reggeized pion exchange, it is necessary to extend this definition to $J=0$ which can be accomplished by analytic continuation of the product in \cref{eq:AJ}.
Using \cref{eq:dJ} with definite negative signature, the limit of $J\to0$ reads
\begin{align} 
\label{eq:pw0}
& 
A^{J\to 0}_{\lambda_\gamma\lambda_i\lambda_f}(s,t)
= \frac{ e_{\pi} \, g_{J\to 0} \, t}{\alpha(t)}\,  (2\lambda_i \lambda_\gamma \delta_{\lambda_i\lambda_f})  \nonumber \\
& \times \sqrt{1-z_t^2} \, 
  \frac{1}{2J} \left[ 
P^{11}_{J-1}(z_t) - 
P^{11}_{J-1}(-z_t)
\right]   \bigg|_{J=0} ~,
\end{align} 
where we have used $d^1_{\lambda_\gamma 0}(\theta_t)=-\lambda_\gamma\sin\theta_t/\sqrt{2}$, and \mbox{$c_0=1$}. 
To explicitly calculate the limit, we use the analytic continuation of the Jacobi polynomials through the hypergeometric function (see \cref{eq:hyp})
\begin{align} \label{eq:hyp11}
P_{J-1}^{11}(z_t) & = J\ \mbox{}_2\tilde F_1\left(1-J,J+2;2;\frac{1-z_t}{2}\right) \ ,
\end{align} yielding
\begin{equation}\label{eq:relJacobiP}
    \frac{1}{2J}(P^{11}_{J-1}(z_t)- P^{11}_{J-1}(-z_t))\bigg|_{J=0}=\frac{2 \, z_t}{z_t^2-1} \ .
\end{equation} 
With this, \cref{eq:pw0} may be written explicitly as
\begin{align} 
\label{eq:pw0-finite}
 A^{J\to 0}_{\lambda_\gamma\lambda_i\lambda_f}(s,t) =   \frac{2 \, e_{\pi} \, g_0 \, t}{-\alpha(t)} (2\lambda_i \lambda_\gamma \delta_{\lambda_i\lambda_f})  \frac{z_t}{\sqrt{1-z_t^2} } \ .
\end{align}
The kinematic pole at $J=0$ in \cref{eq:pw0} which emerges partly from \cref{eq:pwJ} and partly from \cref{eq:dJ} is canceled by a zero at $J=0$ in the Jacobi polynomial in \cref{eq:hyp11}. Therefore, the result of \cref{eq:pw0-finite} is finite, and we may add this contribution to the rest of the summation in \cref{eq:PWexp-t}. 

As $\alpha(t)$ is analytic in $t$, we may expand near $t\sim m_\pi^2$, \mbox{$\alpha(t) = \alpha'(t - m_\pi^2) + O\left((t -m_\pi^2)^2\right)$}, such that \cref{eq:pw0-finite} recovers the simple pion pole and we define the coupling, at $J=0$, to be
\begin{equation}\label{eq:g0}
    g_{J\to 0}=\alpha'g_{\pi NN} \ .
\end{equation}
In the limit of large center-of-mass energy $s$, we have \mbox{$z_t/\sqrt{1-z_t^2}=\cos\theta_t/\sin\theta_t\approx -i$}, leading to:
\begin{align} \label{eq:pw0-pion}
A^{J\to 0}_{\lambda_\gamma\lambda_i\lambda_f}(s,t)
&
\approx -i\, \frac{2 \, e_\pi \, g_{\pi NN} \, t}{m_\pi^2-t} \,(2\lambda_i \lambda_\gamma\delta_{\lambda_i\lambda_f} ) \ .
\end{align}
This expression reminds us of the typical Born amplitude for pion exchange from effective Lagrangians. As will be discussed in the next section, these Born amplitudes recover \cref{eq:pw0-pion} through the mixing of $t$-channel pion and $s$- and $u$-channel nucleon exchange diagrams required by current conservation. In our construction, however, we recovered the current conserving contribution of the $J=0$ pole only considering $t$-channel exchanges. We note that, despite being calculated in the limit $J\to 0$, the angular dependence of \cref{eq:pw0-finite} is nontrivial. This is due to the fact that, strictly speaking, the $J=0$ partial wave vanishes, as $d^0_{\lambda_\gamma,0}(\theta_t) \equiv 0$. The  $A^{J\to 0}_{\lambda_\gamma\lambda_i\lambda_f}(s,t)$ is thus not a partial wave, but rather a new contribution that modifies the other physical partial waves with $J\geq 1$. With an abuse of nomenclature, we keep calling it a partial wave.

\subsection{Spin summation}
\label{sec:spinsum} 
We have demonstrated that the analytic continuation of the amplitude in \cref{eq:vertices_with_reggepole} is finite at $J\to 0$. Therefore, the summation over the $t$-channel angular momentum can include this contribution. Performing this summation, we obtain the Reggeized amplitude for the pion trajectory,
\begin{align}  \label{eq:pionR}
 & A^{\textrm{Regge}}_{\lambda_\gamma\lambda_i\lambda_f}(s,t)
=  A^{J\to0}_{\lambda_\gamma\lambda_i\lambda_f}(s,t) -2 \, e_{\pi}\, t \,  (2\lambda_i\, \lambda_\gamma \, \delta_{\lambda_i\lambda_f})     \nonumber \\ 
&\qquad\times \sum_{J\geq1}(-2)^J g_J \, c_J^2 \, 
\frac{(2J+1) (J+1) }{2J \, (J-\alpha_\pi(t))} 
 \, (p_t \, k_t)^J \nonumber \\
&\qquad \times \frac{1}{2} \left[ 
P^{11}_{J-1}(z_t) - 
P^{11}_{J-1}(-z_t)
\right] \,  \sqrt{1 - z_t^2} \ .  
\end{align}
The kinematical factor $c^2_J$,     
\begin{equation} \label{eq:cJsq}
    c^2_J = \frac{\Gamma^2(J+1) \, e^{J \ln 2}} {\Gamma(2J + 1)} \ ,
\end{equation} 
has alternating single poles for integer $J < 0$, zeros for half-integer $J<0$, and vanishes in the limit $J\to \infty$.

For higher spins, the coupling reflects the internal structure of the hadronic vertices. One expects
 \begin{equation} \label{eq:gJ}
g_J = \alpha' g_{\pi NN} \,
h_J \, (r_t \, r_b)^{J} \ ,
\end{equation} 
where $r_t$ and $r_b$ are related to the hadronic 
radii of interaction in the top and bottom vertices, respectively. The dimensionless constant $h_J$, with $h_0=1$, depends on the details of the microscopic structure of the vertices. 
This parameter can be computed, for example, using quark models, and may, depending on the specifics of the microscopic model, generate singularities in the left half of the complex $J$-plane. 

Taking these analytic properties into account, we approximate \cref{eq:cJsq} by replacing the singularities at $J<0$ by an effective fixed pole at $-j_p$, and the zeros by an effective zero at $-j_z$, while keeping the asymptotic behavior the same:
    \begin{equation}
    c_J^2 \sim \frac{j_p}{j_z} \frac{J+j_z}{J+j_p} \,  e^{-J\log 2} ~.
    \end{equation}
When combined with \cref{eq:gJ}, we have
 \begin{equation}\label{eq:geff} 
 c^2_J \, h_J \, ( r_t\, r_b)^J  \to \frac{j_p}{j_z}\left(
\frac{J+j_z}{J+j_p}\right)  R^{2J} \ , 
\end{equation}
with $j_p\geq 1$, $j_z \geq 1/2$, and $R^2 \equiv   r_t \, r_b/2$ is the effective radius of interaction for the reaction. 

To compare \cref{{eq:pw0-pion},eq:pionR} we remove overall prefactors by defining:
\begin{align}\label{eq:Regge-amp-def}
    A^{\textrm{Regge}}_{\lambda_\gamma\lambda_i\lambda_f}(s,t)&=-i \, 2\,e_{\pi}g_{\pi NN}\, t \, ( 2\lambda_i\lambda_\gamma\delta_{\lambda_i\lambda_f}) \,  \nonumber\\
    &\qquad \times \left[ \mathcal{A}^{J\to0}(s,t) + \mathcal{A}^{J\geq2}_{j_pj_z}(s,t)\right] 
    \ ,
\end{align}
where we refer to the remaining amplitudes, $\mathcal{A}^{J\to0}$ and $\mathcal{A}^{J\geq2}$ as `reduced amplitudes'. They are given by
\begin{equation}
    \label{eq:Pregge0}
    \mathcal{A}^{J\to0}(s,t) = -i \frac{\alpha^\prime}{\alpha(t)} \, \frac{z_t}{\sqrt{1-z_t^2}} ~,
\end{equation}
and 
\begin{align} \label{eq:Pregge}
& \mathcal{A}_{j_pj_z}^{J\geq 2}(s,t)  =-i\alpha' 
\sum_{J\geq 1}\frac{(2J+1)(J+1)}{2J(J-\alpha(t))}\frac{j_p}{j_z}\frac{J+j_z}{J+j_p} \nonumber \\
&\times (-\kappa)^J \frac{1}{2} \left[ 
P^{11}_{J-1}(z_t) - 
P^{11}_{J-1}(-z_t)\right]\, \sqrt{1-z_t^2}\ ,
\end{align}
where $\kappa\equiv p_t \,k_t \, R^2$ is the effective barrier factor.

The summation over $J$ can be readily performed in several ways, \eg using the Sommerfeld-Watson~\cite{Collins:1977jy} transform. Here we use the generating function for the Jacobi polynomials. The partial fraction decomposition of the $J$ dependent prefactors allows us to rewrite
\begin{align} \label{eq:Pregge_eff}
& \mathcal{A}^{J\geq 2}_{j_pj_z}(s,t) = 
-\frac{\alpha'i\kappa }{2}  \, \sqrt{1-z_t^2} \, \Bigg\{
\int_0^1  dy  \bigg[
  \frac{1}{ -\alpha(t) }  \\
 & \qquad + y^{-\alpha(t) } 
 \frac{j_p(\alpha(t)+j_z)(\alpha(t)+1)(2\,\alpha(t)+1)}{\alpha(t) j_z(\alpha(t) + j_p)} \nonumber \\
& \qquad + y^{j_p} \frac{(j_z - j_p)(1-j_p)(1-2j_p)}{j_z(j_p + 
 \alpha(t)) }   \bigg]  \nonumber \\
&\times 2\bigg[  \frac{1}{G(z_t,\kappa y)}\frac{1}{
 (1 + G(z_t,\kappa y))^2 - (\kappa y)^2   }  - (z_t \to -z_t) \bigg] \nonumber \\ 
& +4\,\frac{j_p}{j_z}\bigg[\frac{1}{G(z_t,\kappa)}\frac{1}{(1+G(z_t,\kappa))^2-\kappa^2}-(z_t\to -z_t)\bigg]  \Bigg\} ~. \nonumber
\end{align} 
We have introduced the integral representation \mbox{$(J+a)^{-1}=\int_0^1 dy\, y^{J-1+a}$}, and performed the summation using the generating function of the Jacobi polynomials, which introduces \mbox{$G(z,w)  = \sqrt{1 + 2 w z
 + w^2 }$}. See \cref{app:sum} for details.

The integral in \cref{eq:Pregge_eff} can be computed numerically with fixed values of $R$, $j_p$, and $j_z$. Alternatively, we may consider the dominant contribution of the Regge pole by taking the limit of large $s$ first, and then reducing the dependence on model parameters by taking $j_p,j_z\to\infty$ with the resulting amplitude given by:
\begin{align} \label{eq:Pregge_Jlarge}
 &\mathcal{A}^{J\geq 2}_{j_pj_z}(s,t) \approx 
   \frac{\alpha' }{\alpha(t)} \\
  & \times \left(1  -\tau 
 \frac{2}{\sqrt{\pi}}\Gamma\left(\alpha(t)+\frac{3}{2}\right)\Gamma(1-\alpha(t)) (sR^2)^{\alpha(t)}  \right)\nonumber \\ &\hspace{5cm} + O\left(\frac{1}{s^{j_p}}\right) \nonumber 
\end{align} 
where $\tau=\left[1+\exp(-i\pi\alpha)\right]/2$ is the signature factor. We note that the term $\Gamma(\alpha(t) + 3/2)$ introduces unphysical poles for negative half-integer values of $\alpha(t)$. However, this occurs for values of $t$ beyond the region of validity of \cref{eq:Pregge_larges}. Near $t\sim m_\pi^2$, \cref{eq:Pregge_Jlarge} becomes 
\begin{equation}
    \mathcal{A}^{J\geq 2}(s,m_\pi^2) =-\alpha'\left(2-\frac{i\pi}{2}+\ln\frac{sR^2}{4}\right)+O(m_\pi^2-t)\ ,
\end{equation}
and this contribution is finite because the pion pole is contained in \cref{eq:Pregge0}. The sum of the reduced amplitudes is
\begin{align}\label{eq:Pregge_larges}
    \mathcal{A}^\textrm{Regge}(s,t)&=\mathcal{A}^{J\to0}(s,t) + \mathcal{A}^{J\geq2}_{j_pj_z}(s,t) \\
    &\approx \alpha' \,\tau \,\frac{2}{\sqrt{\pi}} \, \Gamma\bigg(\alpha(t)+\frac32\bigg) \,  \Gamma(-\alpha(t)) \,(sR^2)^{\alpha(t)} \ . \nonumber
\end{align}
At forward angles, $\alpha(0) \sim 0$ and $\Gamma(\alpha + 3/2) \approx \sqrt{\pi}/2$ and \cref{eq:Pregge_larges} resembles the ``Regge propagator'' commonly employed in the literature, such as in Ref.~\cite{Guidal:1997hy}
\begin{align}\label{eq:Pregge-lit}
\mathcal{P}_\pi^{\textrm{Regge}}
    &=\alpha'\,\tau \,\Gamma(-\alpha(t))\left(\frac{s}{s_0}\right)^{\alpha(t)} \ ,
\end{align}
where we can identify $R^{-2}$ with the characteristic scale factor $s_0$. The typical scale of $s_0 = 1\gevsq$  results in an effective interaction radius $R\approx 0.2\fm$ which is much smaller than the expected range of hadronic interactions on the order of $1\fm$. 

Comparing \cref{eq:Pregge0,eq:Pregge_eff,eq:Pregge_Jlarge}, we may systematically investigate several aspects of the pion exchange at high energies. In addition to the overall effect of Reggeization compared to a simple pole amplitude, the effect of subleading terms at finite $s$ can be compared with the $s\to\infty$ limit often assumed in Regge studies. Furthermore, by formally resumming the ladder of higher spin exchanges, the influence of features in the left-half $J$-plane (\ie the effective poles and zeros $j_p,j_z$) may be examined. Finally, because the Reggeized amplitude depends on an energy scale related to the range of interaction, we may compare the conventional choice $R^2=1/s_0$ with more physically motivated values.

\subsection{Results}
\label{sec:results-regge}
\begin{figure}[htbp!]
\begin{center}
  \includegraphics[width=\columnwidth]{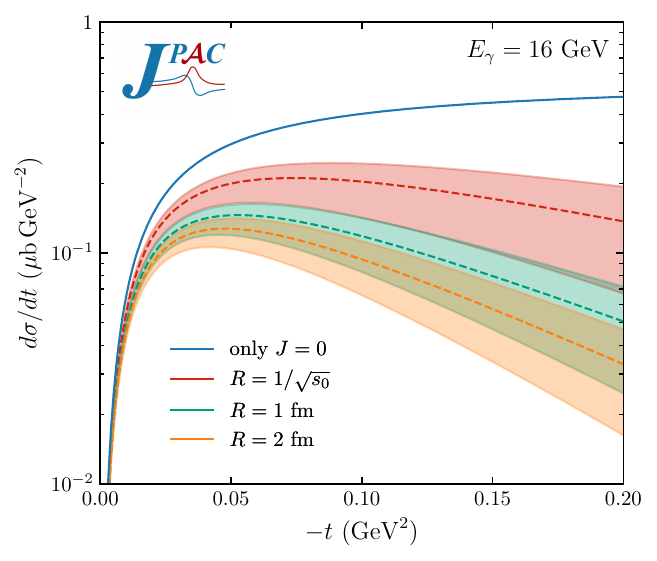}
\end{center}
\caption{Differential cross section computed from the Reggeized amplitude \cref{eq:Regge-amp-def}. 
Different colors of lines and bands correspond to different choices of the interaction radius.
The dashed lines correspond to the Regge pole contribution in the large $s$ limit, as isolated in \cref{eq:Pregge_larges}.
The shaded bands enclose the results of the numerical evaluation of \cref{eq:Pregge_eff} for all possible values of $j_p$ and $j_z$. The upper and lower boundaries correspond to $(j_p,j_z)=(1,\infty)$ and
$(j_p,j_z)=(\infty, \frac{1}{2})$, respectively. 
The $J=0$ contribution only \cref{eq:Pregge0} (solid blue line) is shown for comparison.}
\label{fig:dsigma-regge}
\end{figure}

Now that we have a solid formalism to describe the Reggeization of pion exchange that is gauge invariant by construction, we compute the unpolarized differential cross section,
\begin{equation}\label{eq:cross section}
    \frac{d\sigma}{dt}=\frac{1}{64\pi}\frac{1}{(2m_NE_\gamma)^2}\sum_{\{\lambda\}}|A_{\{\lambda\}}|^2\ ,
\end{equation}
and examine the dependence on the parameters introduced in the description of the Regge couplings. For the pion trajectory we use $\alpha(t)=\alpha'(t-m_\pi^2)$ with $\alpha'=0.7\gevsq$. Our results for a photon beam energy of $E_\gamma=16\gev$ are shown in \cref{fig:dsigma-regge}, in the range $-t\in [0,0.2]\gevsq$. 

We illustrate the results computed using the high-energy limit expression for the reduced amplitude in \cref{eq:Pregge_larges}, for which we took $j_z,j_p\to\infty$ to capture the dominant contribution from the Regge pole, as given by the second line in \cref{eq:Pregge_eff}. 
The uncertainty on this prediction is due to the unknown finite values of $j_p$ and $j_z$. The term in the third line in \cref{eq:Pregge_eff} is subleading to the Regge pole term, and comes with a factor $(j_z-j_p)$ so that values $j_z \gg j_p$ ($j_z \ll j_p$) produce a higher (lower) curve than the single Regge pole (recovered for $j_z = j_p$).

We consider several values for the interaction radius, including the relation with the scale factor $s_0$ required to recover the VGL prescription, \ie $R=1/\sqrt{s_0}\approx 0.2\fm$, as well as physically sound values of $R=1\fm$ and $R=2\fm$. We have checked numerically that the differential cross section obtained with \cref{eq:Pregge-lit} lies on top of that computed using \cref{eq:Pregge_larges} and $R=1/\sqrt{s_0}$ (red dashed). The contribution of the $J=0$ amplitude alone, \ie computed using \cref{eq:Pregge0}, is also shown for comparison (blue solid line). As expected, Reggeization modifies the $t$-dependence of the cross section. The larger the value of $R$, the larger the modification with respect to the $J=0$ only contribution to the cross-section. 

\section{Studies of current conservation in Born-like models}
\label{sec:born}

In the previous section, current conservation was integrated in the covariant structure of the vertex coupling an exchange with spin $J>0$ to the $\gamma\pi$ system. While for $J=0$, \ie the $t$-channel pion exchange, we could not build a current conserving vertex, we showed that the $J=0$ contribution that can be obtained from the $J>0$ result via analytic continuation. The resulting amplitude not only contains the pion pole but also conserves electric current.

In order to analyze the intricacies of pion exchange and understand the origin of the pion pole, it is useful to compare with the framework of Born models for pion photoproduction. In this low-energy approach, amplitudes are built from effective Lagrangians and the pion pole has fixed spin~\cite{Ericson:1988gk}. 

Born models have also been fairly successful in empirically describing $\pi^\pm$ photoproduction data at large values of $s$ and small $t$ since the early work of Cho and Sakurai~\cite{Cho:1969qk}. Extending the low-energy formalism to higher energies often relies on the VGL Reggeization prescription of \cref{eq:prescrip}. A feature of the pion exchange mechanism in pion photoproduction is that it produces a differential cross section that vanishes in the forward direction for kinematical reasons, as seen from \cref{fig:dsigma-regge} for $t\to 0$. This is due to the factor of $t$ in \cref{eq:Regge-amp-def}.  This contradicts experimental results~\cite{Boyarski:1967sp}, where data exhibit a prominent forward peak in the differential cross section. While the narrowness of this peak does suggest a connection to the pion pole, it cannot be explained by a pion exchange alone. In Born models, it is the gauge-invariant combination of the $t$-channel pion exchange and the $s$- (or $u$-) channel nucleon exchange that can describe the observed forward peak. 

In this context, we aim to establish a connection between the Born model and the Reggeization scheme developed in \cref{sec:reggeization}, in order to effectively describe charged pion photoproduction within Regge phenomenology, and at the same time accurately represent the forward behavior of the experimental data.

\subsection{Gauge invariant decomposition of the Born amplitudes}

For the direct channel reaction, \cref{eq:s-channel-reaction}, we write the helicity amplitudes as the contraction of the photon polarization and the hadronic current, 
\begin{equation}\label{eq:amp-schannel}
    A_{\mu_\gamma\mu_i\mu_f}=
    \bar u(p_f,\mu_f)\left(\epsilon_\mu(k,\mu_\gamma) \,J^\mu\right)u(p_{i},\mu_i) \ ,
\end{equation}
and similarly, in the $t$-channel,
\begin{equation}\label{eq:amp-tchannel}
    A_{\lambda_\gamma\lambda_i\lambda_f}=
    \bar u(p_f,\lambda_f)\left(\epsilon_\mu(k,\lambda_\gamma) \, J^\mu\right)v(-p_{i},\lambda_i) \ .
\end{equation} 
The Dirac spinors are given in \cref{eq:spinors-s} and \cref{eq:spinors-t}. 
Note that we use $\mu_x$ for the direct-channel CM frame helicities, while we reserve $\lambda_x$ for the $t$-channel helicities as in the previous section.
In either case, the current $J^\mu$ must couple to all asymptotic states in all $s$-, $t$-, and $u$-channels, which suggests that the amplitude contains both nucleon and pion poles that cannot be separated in a gauge-invariant way. The presence of simultaneous poles in $s$-, $t$- and $u$- channels is essential when it comes to the Reggeization of $t$-channel exchanges. 

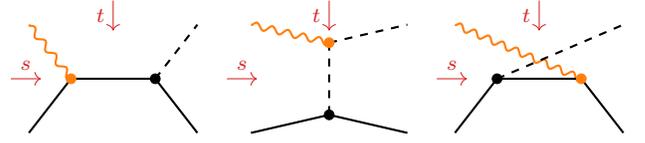
\begin{figure}[t]
\begin{subfigure}[b]{0.33\columnwidth}
\centering
  \begin{tikzpicture}[feynman scale=0.8, baseline=(l1)]
    \begin{feynman}[small]
      \vertex (v1) at (-0.7,0);
      \vertex (v2) at (0.7,0);
      \vertex (l1) at (-1.4,0.9) ;
      \vertex (l2) at (-1.4,-0.9) ;
      \vertex (r1) at (1.4,0.9) ;
       \vertex (r2) at (1.4,-0.9) ;
      \diagram*{
        (l1) -- [jpac-orange, photon, thick] (v1), 
        (v2) -- [scalar, thick] (r1),
        (l2) -- [plain, thick] (v1),
        (v2) -- [plain, thick] (r2),
        (v1) -- [plain, thick] (v2)
        };
       \draw[blob, jpac-orange] (v1) circle(0.08) ;
       \draw[blob, black] (v2) circle(0.08) ;
       \end{feynman}
        \draw [jpac-red,->] (-1.7,0) -- (-1.2,0) node [pos=0.5,above,font=\footnotesize] {\(s\)};
        \draw [jpac-red,->] (0,1.3) -- (0,0.8) node [pos=0.5,left,font=\footnotesize] {\(t\)};
  \end{tikzpicture}
\end{subfigure}
\begin{subfigure}[b]{0.3\columnwidth}
\centering
\begin{tikzpicture}[feynman scale=0.8, baseline=(l1)]
    \begin{feynman}[small]
      \vertex (v1) at (0,0.6);
      \vertex (v2) at (0,-0.6);
      \vertex (l1) at (-1.3,0.9) ;
      \vertex (l2) at (-1.3,-0.9) ;
      \vertex (r1) at (1.3,0.9) ;
      \vertex (r2) at (1.3,-0.9) ;
      \diagram*{
        (l1) -- [jpac-orange, photon, thick] (v1), 
        (l2) -- [plain, thick] (v2),
        (v1) -- [scalar, thick] (r1),
        (v2) -- [plain, thick] (r2),
        (v1) -- [scalar, thick] (v2)
       };
        \draw[blob, jpac-orange] (v1) circle(0.08) ;
       \draw[blob, black] (v2) circle(0.08) ;
       \end{feynman}
        \draw [jpac-red,->] (-1.7,0) -- (-1.2,0) node [pos=0.5,above,font=\footnotesize] {\(s\)};
        \draw [jpac-red,->] (0,1.3) -- (0,0.8) node [pos=0.5,left,font=\footnotesize] {\(t\)};
  \end{tikzpicture}
\end{subfigure}
\begin{subfigure}[b]{0.33\columnwidth}
\centering
  \begin{tikzpicture}[feynman scale=0.8, baseline=(l1)]
    \begin{feynman}[small]
      \vertex (v1) at (-0.7,0);
      \vertex (v2) at (0.7,0);
      \vertex (l1) at (-1.4,0.9) ;
      \vertex (l2) at (-1.4,-0.9) ;
      \vertex (r1) at (1.4,0.9) ;
       \vertex (r2) at (1.4,-0.9) ;
      \diagram*{
        (l1) -- [jpac-orange, photon, thick] (v2), 
        (v1) -- [scalar, thick] (r1),
        (l2) -- [plain, thick] (v1),
        (v2) -- [plain, thick] (r2),
        (v1) -- [plain, thick] (v2)
        };
        \draw[blob, black] (v1) circle(0.08) ;
       \draw[blob, jpac-orange] (v2) circle(0.08) ;
       \end{feynman}
        \draw [jpac-red,->] (-1.7,0) -- (-1.2,0) node [pos=0.5,above,font=\footnotesize] {\(s\)};
        \draw [jpac-red,->] (0,1.3) -- (0,0.8) node [pos=0.5,left,font=\footnotesize] {\(t\)};
  \end{tikzpicture}
\end{subfigure}
\caption{Feynman diagrams of the $s$-, $t$-, and $u$-channel Born terms contributing to charged pion photoproduction. The arrows indicate the time direction for the $s$-channel reaction, \cref{eq:s-channel-reaction}, and the $t$-channel reaction, \cref{eq:t-channel-reaction}.}
\label{fig:BornDiagrams}
\end{figure}

In the Born model, the hadronic current can be generically written as
$J^\mu=J^{\mu}_{\pi}+J^{\mu}_{N_i}+J^{\mu}_{N_f}$, where the contributions of each of the Born diagrams depicted in \cref{fig:BornDiagrams} are given in several textbooks (see \eg Ref.~\cite{Ericson:1988gk}),
\begin{subequations} \label{eq:born-currents}
\begin{align} \label{eq:born-pion}
& J^{\mu}_{\pi} = -\sqrt{2} \,e_{\pi} \, g_{\pi NN} \,\frac{ q_{\pi}^\mu - p_{\pi}^\mu}{t-m_\pi^2}   \, \gamma_5   \, ,\\
& J^{\mu}_{N_i} = \sqrt{2} \, e_{ N_i} \,g_{\pi NN} \, \gamma_5  \frac{\slashed{q}_{{N_i}} +m_N}{s-m_N^2} 
   \,  \gamma^\mu  \, ,
   \\
 & J^{\mu}_{N_f} = \sqrt{2} \, e_{N_f} \, g_{\pi NN}  \,\gamma^\mu \frac{\slashed{q}_{{N_f}} +m_N}{u-m_N^2} 
    \, \gamma_5  \, ,
\end{align}
\end{subequations}
The four-momenta carried by the exchanged particles are defined as \mbox{$q_{N_i}=k+p_i$}, \mbox{$q_{\pi}=k-p_\pi$}, and \mbox{$q_{N_f}=p_f-k$ ($q_{N_i}^2=s$}, \mbox{$q_{\pi}^2=t$}, and \mbox{$q_{N_f}^2=u$}). 
We label the amplitudes by the electric charge of the external particles, \ie \mbox{$e_\pi\to\{e_{\pi^+}=+e,\,e_{\pi^-}=-e\}$}, and \mbox{$e_{N_{i,f}}\to\{e_p=+e,\,e_n=0\}$}. For $\pi^+$ ($\pi^-$) photoproduction off protons (neutrons), the contribution to the current from the $u$- ($s$-)channel diagram exactly vanishes. The pseudoscalar coupling of the pion to the nucleons is $g_{\pi NN} = 13.48$~\cite{Matsinos:2019kqi}.\footnote{There exist two different descriptions of meson-baryon interactions in photoproduction frequently used in the literature: pseudoscalar and pseudovector couplings. When using pseudovector coupling, one needs to additionally consider the Kroll-Ruderman contact diagram. In the case of pion photoproduction, it is straightforward to show the equivalence between the two coupling schemes, given that nucleons satisfy the free Dirac equation, $(\slashed{p}-m_N)u(p,\mu)=0$. The two coupling constants are related via $g_{\pi NN}/2m_N=f_{\pi NN}/m_\pi$, where $f_{\pi NN}$ is the pseudovector coupling constant ($f_{\pi NN}^2/4\pi=0.08$).} The factors $\sqrt{2}$ arise from the relation between the charge and the isospin amplitudes. 

The individual Feynman diagrams generate the Lorentz covariant currents in \cref{eq:born-currents}. However, after contracting with the photon polarization vector, each resulting component is not Lorentz invariant. This is because real photons have two physical polarization states and their polarization vectors are frame dependent. It is easy to see, for instance, that the pion-exchange amplitude given by the current in \cref{eq:born-pion} is not gauge invariant: it fails to satisfy the Ward identity, $k_\mu J^{\mu}_\pi\neq 0$, whereas the total current does satisfy it, $k_\mu J^\mu=0$, since the charges must obey the relation $e_{N_i}=e_\pi+e_{N_f}$ for charge conservation. This is a well-known result in Born-like models: since both pions and nucleons carry electric charge, local phase invariance requires that both couple to the photon, hence both $s$- ($u$-) and $t$-channel diagrams are needed for $\pi^+$($\pi^-$) production.

The fact that the individual channels are not gauge invariant suggests that the splitting of the production mechanism in Born diagrams is somewhat arbitrary. 
Here, we argue that it is more desirable to identify the piece of the nucleon exchange term that is necessary and sufficient to restore gauge invariance of the pion exchange term. 
As a first step, it is convenient to separate the nucleon-exchange amplitudes in \cref{eq:amp-schannel}, with the currents in \cref{eq:born-currents}, into electric and magnetic components. The magnetic amplitude \mbox{$A^{\textrm{m}}_{\mu_\gamma\mu_i\mu_f}$} is defined as being proportional to  \mbox{$\sigma^{\mu\nu}k_\nu=\frac{i}{2}[\gamma^\mu,\gamma^\nu]k_\nu$}, and the electric one \mbox{$A^{\textrm{e}}_{\mu_\gamma\mu_i\mu_f} = A_{\mu_\gamma\mu_i\mu_f} - A^{\textrm{m}}_{\mu_\gamma\mu_i\mu_f}$} contains the remainder. 
The contribution of the anomalous magnetic term, which is gauge-invariant per se, is negligible and not included. Explicitly,
\begin{widetext}
\begin{subequations}\label{eq:amp-el-mag}
    \begin{align}\label{eq:amp-electric}
        A^{\textrm{e}}_{\mu_\gamma\mu_i\mu_f}&=\sqrt{2} \, g_{\pi NN}\,\epsilon_\mu(k,\mu_\gamma)\left[e_\pi\frac{2p_\pi^\mu-k^\mu}{t-m_\pi^2}  +e_{N_i}\frac{2p_i^\mu+k^\mu}{s-m_N^2}+e_{N_f}\frac{2p_f^\mu-k^\mu}{u-m_N^2}\right]\bar{u}(p_f,\mu_f)\,\gamma_5\,u(p_i,\mu_i) \ , \\ \label{eq:amp-magnetic}
        A^{\textrm{m}}_{\mu_\gamma\mu_i\mu_f}&=\sqrt{2} \, g_{\pi NN}\left[ \frac{e_{N_i}}{s-m_N^2}+\frac{e_{N_f}}{u-m_N^2}\right]\bar{u}(p_f,\mu_f)\,\gamma_5 \, i\sigma^{\mu\nu} \, k_\mu \, \epsilon_\nu(k,\mu_\gamma)\,u(p_i,\mu_i) \ ,
    \end{align}
\end{subequations}
\end{widetext}
Although this separation between the electric and magnetic components of the nucleon Born amplitude seems natural, it is not commonly performed in the literature. Instead, the full Born amplitude is often inaccurately referred to as ``electric Born amplitude'', as opposed to the anomalous magnetic amplitude. It is important to keep this distinction in mind when comparing the results presented below with those in the literature, for example, with those in Ref.~\cite{Guidal:1997hy}.

The two amplitudes in \cref{eq:amp-el-mag} are gauge invariant by themselves.
Since the pion has no magnetic moment, it contributes only to the electric current. This suggests that the Reggeization of the pion should only involve the electric components of the nucleon exchanges. On the other hand, the magnetic term contributes at small momentum transfer $-t$, where the electric amplitude vanishes. As a result, the magnetic part of the nucleon Born term can explain the forward structure observed in the cross section data in charged pion photoproduction experiments. The contribution of the magnetic Born amplitude is further discussed later in \cref{sec:magneticterm}. 

Gauge invariant amplitudes can be re-expressed in terms of the (gauge invariant) electromagnetic field tensor,
\begin{equation}
    F_{\mu\nu}(k,\mu_\gamma)=\epsilon_\mu(k,\mu_\gamma) \,k_\nu-  k_\mu \, \epsilon_{\nu}(k,\mu_\gamma) \ ,
\end{equation}
as

\begin{equation}\label{eq:mgi-Lorentz}
   F_{\mu\nu} \, P^\mu \,p^\nu_{\pi}= (\epsilon\cdot P)\left(k\cdot p_\pi\right) -\left(\epsilon\cdot p_\pi \right)(k\cdot P) \ .
\end{equation}
We introduced the simplified notation $\epsilon_\mu$ for the photon polarization vector.
Using $t-m_\pi^2=k^2-2(k\cdot p_\pi)$ and $s-u=2k\cdot(p_i+p_f)=2(k\cdot P)$, we can then write an expression for the ``minimal gauge invariant'' (MGI) pion exchange amplitude,
\begin{align}\label{eq:mgi}
    A^{\pi \textrm{-MGI}}_{\mu_\gamma\mu_i\mu_f}&=2\sqrt{2}e_\pi g_{\pi NN} \left(\frac{\epsilon\cdot (p_\pi - k/2)}{t-m_\pi^2}+\frac{\epsilon\cdot P}{s-u}\right)\nonumber \\
    &\times \bar{u}(p_f,\mu_f)\gamma_5u(p_i,\mu_i) \ .
\end{align}
The addition of the second term in \cref{eq:mgi}, which originates from contributions from the nucleon Born diagrams, restores the gauge invariance of the ``bare'' pion exchange given by the first term.
Indeed, by using both energy-momentum conservation ($k+p_i=p_\pi+p_f$) and charge conservation ($e_\pi=e_{N_i}-e_{N_f}$), the electric amplitude in \cref{eq:amp-electric} can be written as a sum of three terms, each individually gauge invariant and proportional to the electric charge of the exchanged particle,  
\begin{widetext}
\begin{align}\label{eq:electric-3terms}
    &A^{\textrm{e}}_{\mu_\gamma\mu_i\mu_f}=2\sqrt{2}g_{\pi NN}\bigg[  e_\pi\left(\frac{\epsilon\cdot (p_\pi-k/2)}{t-m_\pi^2}+\frac{\epsilon\cdot P}{s-u}\right)  \\
     &+\frac12e_{N_i}\left(\frac{\epsilon\cdot p_\pi}{s-m_N^2}+\frac{\epsilon\cdot P}{s-u}\frac{t-m_\pi^2-k^2}{s-m_N^2}\right) -\frac12e_{N_f}\left(\frac{\epsilon\cdot p_\pi}{u-m_N^2}+\frac{\epsilon\cdot P}{s-u}\frac{t-m_\pi^2-k^2}{u-m_N^2}\right)\bigg]\bar{u}(p_f,\mu_f)\gamma_5u(p_i,\mu_i) \ . \nonumber
\end{align}
\end{widetext}
Now that we checked explicitly gauge invariance, we can restrict to real photons enforcing $\epsilon \cdot k = k^2 =0$.

\subsection{Electric Born terms}

Now we turn to the discussion of the frame dependence of the pion Born term, and of the various contributions to the full Born amplitude in general, that urged the construction of the Reggeization scheme in \cref{sec:reggeization} and the decomposition of the full Born amplitude in the gauge invariant pieces in \cref{eq:electric-3terms}. While the detailed calculation of the amplitudes, as well as definitions of the kinematic variables and the nucleon spinors, are given in \cref{app:Born-s,app:Born-t} for the $s$- and $t$-channel frames, respectively, here we summarize the main observations.

The contribution from the bare pion-exchange, \ie that in \cref{eq:born-pion}, is proportional to the product of the photon polarization vector and the pion four-momentum, $(\epsilon\cdot p_\pi)$. In the $s$-channel frame we have $(\epsilon\cdot p_\pi)_s\sim q_s\sin\theta_s$, which goes as $\sqrt{-t}$ in the large $s$ limit, while the terms related to the nucleon electric amplitude are suppressed by a factor of $s$.

However, when evaluated in the $t$-channel frame, the bare pion-exchange term vanishes exactly, $(\epsilon\cdot p_\pi)_t=0$, because $p_\pi$ is antiparallel to the photon four-momentum $k$, making it perpendicular to the purely spacelike $\epsilon(k,\lambda_\gamma)$. This indicates that, for the Born approach to be compatible with Regge phenomenology, the partial waves must incorporate information about the $s$- ($u$-)channel nucleon term. 

The leading term in the $t$-channel frame is precisely the piece of the electric nucleon Born terms that we required in \cref{eq:mgi} to make the pion exchange gauge invariant. This term is proportional to $(\epsilon\cdot P)$, which in the $s$-channel frame is $(\epsilon\cdot P)_s\sim q_s\sin\theta_s$. Therefore, this term is subdominant in the $s$-channel frame, because of the additional denominator $s-u$. But in the $t$-channel frame this is $(\epsilon\cdot P)_t$, which behaves as $p_t\sin\theta_t\sim i \,s\sqrt{t}/(m_\pi^2-t)$. 

In this limit, the electric Born amplitude in the $t$-channel frame reads
\begin{equation}   \label{eq:slimit-tchannel-v2}
A^\textrm{e}_{\lambda_\gamma \lambda_i\lambda_f}(s,t)\approx -i\, \frac{2\,e_\pi\, g_{\pi NN} \, t}{m_\pi^2-t} \,\left(2\lambda_i {\lambda_\gamma}\delta_{\lambda_i\lambda_f}\right)  \ ,
\end{equation}
which is precisely the $t$-channel partial wave amplitude with $J=0$ [\cf \cref{eq:Regge-amp-def,eq:Pregge0}]. 

Let us take a different approach to discussing the frame dependence of the Born terms. To further exploit the connection between pion and nucleon exchanges, it is instructive to study the individual contributions to the physical observables, \eg the total cross section, calculated as the amplitude squared summed over physical polarizations,  $\overline{A}^{\,2}\equiv\frac{1}{4}\sum_{\{\mu\}}|A_{\{\mu\}}|^2$, with $\frac{d\sigma}{dt}\propto\overline{A}^{\,2}$.
 We focus on $\pi^+$ photoproduction,  $\pi^-$ being analogous.

An amplitude that breaks gauge invariance leads to a cross section that depends on the reference frame. This is the case when Born terms are taken individually. If we calculate the cross section from the (gauge-invariant) electric amplitude as decomposed in \cref{eq:amp-electric} where each term is not individually gauge invariant, and choose the $s$-channel frame for the calculation, the contribution of the $s$-channel proton exchange vanishes due to $(\epsilon\cdot p_i)_s=0$, and for the $u$-channel neutron exchange it vanishes because $e_{N_f}=0$. Therefore, the cross section must come entirely from the pion exchange, \ie from the first term, and we have
\begin{align}\label{eq:amp-electric-piplus}
    \overline{A}_\textrm{e}^2&=\overline{A}_{\textrm{Born-}\pi , s}^2 \ .
\end{align}
We may compare this with the square of the MGI amplitude in \cref{eq:mgi}, which includes the same pion contribution but with minimal additional terms to satisfy gauge invariance. The resulting cross sections can be related by,
\begin{equation}\label{eq:mgi-pion}
    \overline{A}^{\,2}_{\pi\textrm{-MGI}}=F^2\,\overline{A}^{\,2}_{\textrm{Born-}\pi , s} \ ,
\end{equation}
where
\begin{equation}
 F=2\,\,\frac{s-m_N^2}{s-u} \ . 
\end{equation}
At large $s$, \ie in the Regge limit,  $s\gg m_N^2$ and $t\simeq 0$, so that $F\approx 1$. Thus when working entirely in the $s$-channel frame, the differential cross-section can be well approximated by solely the pion Born term of \cref{eq:amp-electric} despite it not being gauge invariant.  

We now repeat the calculations in the $t$-channel frame, where the situation is completely different, as the only nonvanishing contribution originates from the proton exchange diagram:
\begin{align}\label{eq:amp-electric-piplus-t-chan}
    \overline{A}_\textrm{e}^2&=\overline{A}_{\textrm{Born-}p, t}^2 \ , 
\end{align}
and comparing with \cref{eq:amp-electric-piplus}, we must conclude   $\overline{A}^{\,2}_{\textrm{Born-}p, t} = \overline{A}^{\,2}_{\textrm{Born-}\pi, s}$. 
This suggests that, when viewed from the $s$-channel frame, the pion gives the full cross section, while in the $t$-channel it must originate from the nucleon.

Thus the MGI amplitude in \cref{eq:mgi-pion} must actually incorporate both nucleon and pion exchange components, to always generate the pion pole regardless of the reference frame. 
As this is a consequence of gauge invariance, it suggests an important question: How does an amplitude arising from nucleon exchange diagrams, which is a priori agnostic of any information regarding the pion, generate a pion pole when evaluated in a different frame? We consider the expression for the photon polarization vector when summed over physical helicities:
\begin{align}\label{eq:sum-pol}
    \overline{\epsilon}_{\mu\nu}&\equiv \sum_{\mu_\gamma=\pm 1}\epsilon_{\mu}(k,\mu_\gamma) \, \epsilon^*_{\nu}(k,\mu_\gamma) \nonumber \\ &=-g_{\mu\nu}-\frac{k_\mu k_\nu}{(k\cdot n)^2}+\frac{k_\mu n_\nu + k_\nu n_\mu}{k\cdot n} \ ,
\end{align}
where $n=(1,\mathbf{0})$ is a fixed time-like four-vector.\footnote{Note because $n$ is purely time-like it is invariant under spatial rotations and \cref{eq:sum-pol} is valid regardless of the direction of motion of the photon.} The contribution to the average squared amplitude of a given Born exchange is then obtained by contracting \cref{eq:sum-pol} with the corresponding momenta. For the nucleon exchanges, we have an overall factor of
\begin{align}\label{eq:sum-pol-momenta}
    \overline{\epsilon}_{\mu\nu} \, P^\mu P^\nu& \propto \frac{1}{(k\cdot n)^2} \ .
\end{align}
Evaluating this in the two different frames, we see
\begin{subequations}
    \begin{align}
        (k\cdot n)_s&=E_\gamma^s=\frac{s-m_N^2}{2\sqrt{s}} \ , \\
        (k\cdot n)_t&=E_\gamma^t=\frac{t-m_\pi^2}{2\sqrt{t}} \ , 
    \end{align}
\end{subequations}
which gives rise to a nucleon pole in the $s$-channel frame, but it is responsible for a pion pole when evaluated in the $t$-channel frame. 

\subsection{Magnetic term}
\label{sec:magneticterm}
Now we come to the contribution to the amplitude from the magnetic term of the nucleon Born diagrams. As previously discussed, the magnetic piece of the nucleon amplitude in \cref{eq:amp-magnetic} is gauge invariant on its own. 
The expression of the magnetic amplitude has little dependence on $t$, and it becomes a constant at large $s$.
Therefore, the magnetic amplitude contributes at small momentum transfer, where the electric amplitude vanishes, and is needed to explain the forward structure observed in the cross section data in charged pion photoproduction experiments. Alternative explanations of the forward peaking data proposed within the framework of Regge theory included absorption or Regge cut models (see \eg Refs.~\cite{Henyey:1969qe,Williams:1970rg}) and models based on the existence of a parity doublet of the pion (see \eg Refs.~\cite{Ball:1968zza,Henyey:1968zza}). However, the main focus of the present work was to establish a suitable Reggeization scheme for pion exchange (see~\cref{sec:reggeization}), and we will not extend the discussion on the origin of the forward peak in the cross section.

\section{Numerical results}
\label{sec:results}

\begin{figure}[htbp!]
\begin{center}
  \includegraphics[width=\columnwidth]{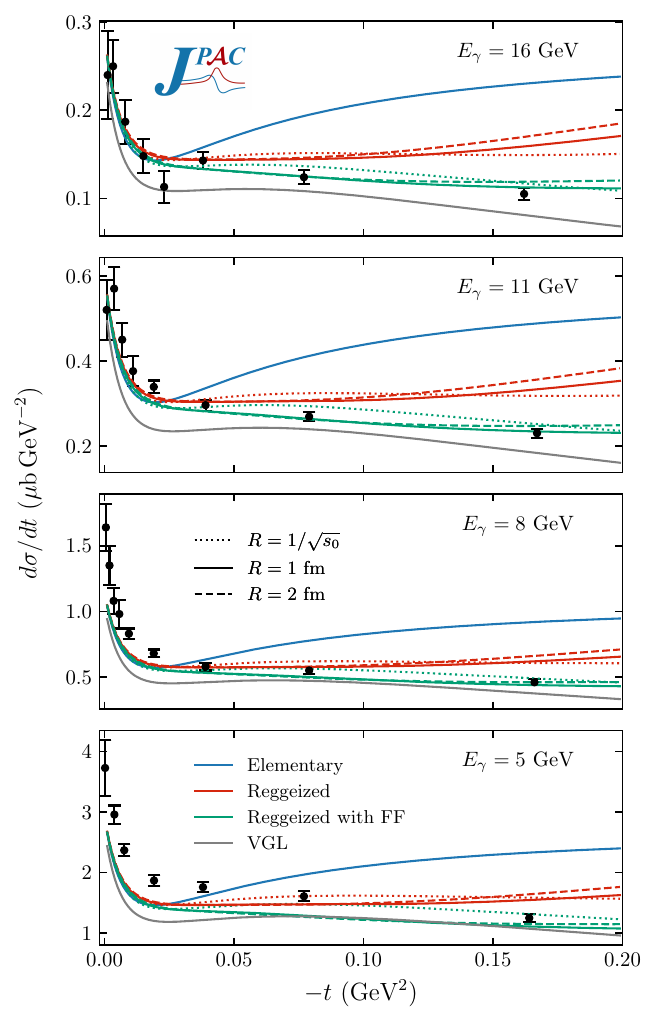}
\end{center}
\caption{Differential cross section computed from Reggeized electric amplitudes with the addition of the magnetic amplitudes without and with a form factor, as described in the text. The elementary exchange Born amplitudes and the Reggeization prescription of VGL~\cite{Guidal:1997hy} are shown for comparison.
 The experimental data is taken from Ref.~\cite{Boyarski:1967sp}. 
}
\label{fig:dsigma-regge-total}
\end{figure}

Now that we have a solid understanding of the role of gauge invariance in both Reggeization of the pion as well as in Born models, we present results for the total cross sections. Our results for $\pi^+$ photoproduction are shown in \cref{fig:dsigma-regge-total} for different values of the beam energy. 

The nucleon magnetic amplitude is approximately constant in $t$. While it is needed to reproduce the forward peak, it provides too large of a contribution at finite $t$. For this reason, we suppress the magnetic amplitude with a dipole form factor $\beta(t) = \Lambda^2/(\Lambda^2 -t)$, with $\Lambda\sim 1\gevsq$. 
We also include the remainder of the electric contribution of the $s$-channel nucleon exchange, although we have shown in \cref{sec:born} that it is negligible at high energies. The experimental data from Ref.~\cite{Boyarski:1967sp} are displayed for comparison. We also compare with the results from elementary exchange Born amplitudes, as well as with the VGL Reggeization prescription~\cite{Guidal:1997hy}.\footnote{The VGL curve is calculated out of the equations presented in~\cite{Guidal:1997hy}, but differs slightly from the one plotted in Fig.~5 of the same reference.}

For $-t\lesssim 0.05$--$0.1\gevsq$, all models considered reproduce the experimental cross section similarly well, particularly at the highest energies. However, at larger values of $t$, the models without the form factor are unable to describe the data. This can be explained by the fact that as the momentum transfer increased, the nucleon magnetic amplitudes dominate over the increasingly smaller Regge amplitudes. Therefore, the use of a form factor on the magnetic amplitudes enables the modulation of the $t$-dependence of the differential cross section. This approach is able to reproduce the data in the range $-t\lesssim 0.2\gevsq$. While the bare pion cannot accommodate an opposite-helicity interaction in the $N\bar{N}$ vertex, the Reggeized pion in principle can. This interaction would contribute at large energies with extra factors of $\sqrt{t}$ which can help fill in the discrepancy at larger $t$. 
Additionally, other Regge trajectories, such as natural parity exchanges, come into play.

We note that the results of the VGL prescription to Reggeize the full Born amplitude systematically fall below the data at all energies. 

\section{Summary and Conclusions}
\label{sec:conclusions}

In this work, we have presented an approach to Reggeize pion exchange in charged pion photoproduction reactions. 
Historically, a great deal of theoretical work has been dedicated to explaining experimental data in this and related reactions, in particular in the region at high energies and small momentum transfer where Regge pion exchange dominates. Our approach reconciles gauge invariance with analyticity in the complex angular momentum plane. 

The method involves explicitly summing over $t$-channel helicity partial waves with even spin, which we have built from covariant vertices. As a result, the partial-wave amplitudes are gauge invariant by construction for $J\geq 2$. The key is that the analytic continuation to $J=0$ produces a gauge-invariant amplitude that can be identified as the leading contribution of the electric Born amplitude at high energies. This naturally solves the problem posed by a vanishing pion exchange amplitude when evaluated in the $t$-channel frame. Based on phenomenological grounds, we have argued that the $J=0$ pole in the $t$-channel can be interpreted as a component of the electric nucleon-exchange amplitude. This extra component is needed to restore the gauge invariance of the pion-exchange amplitude. Therefore, the conspiracy between the pion and the nucleon Born diagrams not only explains the very forward peak observed in the cross section but also enables a consistent approach to pion Reggeization.
By making physically sound assumptions about the Regge couplings, we have been able to perform the summation over $t$-channel partial waves, including the new $J=0$ term, and obtain an integral expression for the Regge amplitude. At high energies and considering only the dominant Regge pole contribution, this expression simplifies to an algebraic form, providing a more justified expression compared to other Reggeization prescriptions employed in the literature.

This work serves as a stepping stone toward understanding high-energy meson-exchange reactions. Although we have considered charge pion photoproduction for its simplicity, the formalism presented can be readily applied to more complicated reactions. For instance, the same intricacies of gauge invariance and pion Reggeization are also present in the photoproduction of $\pi^-\Delta^{++}$~\cite{JointPhysicsAnalysisCenter:2017del}. Comparing with analogous production mechanisms in other reactions such as elastic charge-exchange $\pi N$ scattering would also allow us to better understand peripheral hadron-hadron interactions. The extension to kaon exchanges is also relevant to interpret strange production at GlueX~\cite{GlueX:2021pcl,Pauli:2022ehd}.   

\acknowledgments

This work was supported by the U.S. Department of Energy contract DE-AC05-06OR23177, under which Jefferson Science Associates, LLC operates Jefferson Lab 
and also by the U.S. Department of Energy Grant 
Nos.~\mbox{DE-FG02-87ER40365}, 
and it contributes to the aims of the U.S. Department of Energy \mbox{ExoHad} Topical Collaboration, contract \mbox{DE-SC0023598}.
CFR is supported by Spanish Ministerio de Ciencia, Innovación y Universidades \mbox{(MICIU)} under Grant \mbox{No.~BG20/00133}.
NH, VM, and RJP have been supported by the projects \mbox{CEX2019-000918-M} (Unidad de Excelencia ``María de Maeztu''), \mbox{PID2020-118758GB-I00}, financed by \mbox{MICIU/AEI/10.13039/501100011033/} and FEDER, UE, as well as by the EU \mbox{STRONG-2020} project, under the program \mbox{H2020-INFRAIA-2018-1} Grant Agreement \mbox{No.~824093}. 
VM is a Serra H\'unter fellow and acknowledges support from \mbox{CNS2022-136085}. 
VS and WAS acknowledge the support of the U.S.~Department of Energy ExoHad Topical Collaboration, contract DE-SC0023598.

\appendix

\section{Born amplitudes in the $s$-channel}\label{app:Born-s}
In the $s$-channel frame, the photon momentum is taken along the $+\mathbf{\hat{z}}$ direction, and the pion and the recoiling nucleon are contained in the $xz$ plane. The four momenta of the particles are
\begin{subequations}
    \begin{align}
        k^\mu&=(k_s,0,0,k_s) \ , \\
        p_i^\mu&=(E_i^s,0,0,-k_s) \ , \\
        p_\pi^\mu&=(E_\pi^s,q_s\sin\theta_s,0,q_s\cos\theta_s) \ , \\
        p_f^\mu&=(E_f^s,-q_s\sin\theta_s,0,-q_s\cos\theta_s) \ .
    \end{align}
\end{subequations}
The polarization vector of the photon is given by
\begin{equation}\label{eq:polvec_photon}
     \epsilon(k,\mu_\gamma)=\frac{1}{\sqrt{2}}\left(0,- \mu_\gamma,-i,0\right) \ ,
\end{equation}
where $\mu_\gamma=\pm 1$ is the photon helicity. The helicities of the nucleons are $\mu_i=\pm 1/2$ and $\mu_f=\pm 1/2$.

With these definitions, the electric amplitude as written in Eq.~(\ref{eq:electric-3terms}) reads:
\begin{align}
        A^{\textrm{e}}_{\mu_\gamma\mu_i\mu_f}&=2g_{\pi NN}{\mu_\gamma q_s\sin\theta_s}\bigg[  e_\pi\left(\frac{1}{t-m_\pi^2}-\frac{1}{s-u}\right) \nonumber \\
     &+\frac12 e_{N_i}\left(\frac{1}{s-m_N^2}-\frac{1}{s-u}\frac{t-m_\pi^2}{s-m_N^2}\right)  \nonumber \\
     &-\frac12 e_{N_f}\left(\frac{1}{u-m_N^2}-\frac{1}{s-u}\frac{t-m_\pi^2}{u-m_N^2}\right)\bigg] \nonumber \\ 
     &\times\,\bar{u}(p_f,\mu_f)\,\gamma_5\, u(p_i,\mu_i) \ . 
\end{align}

The photon energy $k_s$, the nucleon energies $E_i^s$ and $E_f^s$, and the pion energy $E_\pi^s$ and momentum $q_s$ are
\begin{subequations}
    \begin{align}
        k_s&=(s-m_\pi^2)/2\sqrt{s} \ , \\
        E_i^s&=(s+m_N^2)/2\sqrt{s} \ , \\
        E_f^s&=(s+m_N^2-m_\pi^2)/2\sqrt{s} \ , \\
        E_\pi^s&=(s-m_N^2+m_\pi^2)/2\sqrt{s} \ , \\
        q_s&=\frac{\sqrt{\left(s-(m_N+m_\pi)^2\right)\left(s-(m_N-m_\pi)^2\right)}}{2\sqrt{s}} \ ,
    \end{align}
\end{subequations}
The cosine and sine of the scattering angle in the center of mass of the $s$-channel, $\theta_s$, read
\begin{subequations}
    \begin{align}
        \cos\theta_s&=\frac{t-u+\Delta/s}{4k_sq_s} \ , \\
        \sin\theta_s&=\frac{\sqrt{\phi/s}}{2k_sq_s} \ ,
    \end{align}
\end{subequations}
with $\Delta=m_N^2(m_N^2-m_\pi^2)$ and the Kibble function \mbox{$\phi=stu-m_\pi^2m_N^2(m_\pi^2-t)-tm_N^4>0$}.

In the limit $s\to\infty$, it follows that $q_s\sin\theta_s\approx \sqrt{-t}$ and $-u\approx s-2m_N^2$, and it is easy to see that the terms in the electric part of the amplitude that come from the nucleon exchange are suppressed by a factor of $s$ with respect to the pion exchange:
\begin{align}
        A^{\textrm{e}}_{\mu_\gamma\mu_i\mu_f}&\approx 2g_{\pi NN}{\mu_\gamma\sqrt{-t}}\bigg[  e_\pi\left(\frac{1}{t-m_\pi^2}-\frac{1}{2(s-m_N^2)}\right) \nonumber \\
     &+\frac12 e_{N_i}\left(\frac{1}{s-m_N^2}-\frac{t-m_\pi^2}{2(s-m_N^2)^2}\right) \nonumber \\
     &+\frac12 e_{N_f}\left(\frac{1}{s-m_N^2}-\frac{t-m_\pi^2}{2(s-m_N^2)^2}\right)\bigg]\nonumber \\
     & \times\,\bar{u}(p_f,\mu_f)\,\gamma_5\, u(p_i,\mu_i) \ . 
\end{align}
The symbol $\approx$ stands for the leading behavior at large $s$. 

The initial and final nucleons move along the directions $-\mathbf{\hat{k}}_s$ and $-\mathbf{\hat{q}}_s$, respectively. We use the second-particle phase convention of Jacob and Wick~\cite{Jacob:1959at}. Therefore, the Dirac spinors that we use are
\begin{subequations}\label{eq:spinors-s}
    \begin{align}
        u(p_i,\pm1/2)&=\beta_i^s\begin{pmatrix}
            \chi_\mp(0) \\ \pm\alpha_i^s \chi_\mp(0)
        \end{pmatrix} \equiv u_{\pm}\ , \\
        \bar u(p_f,\pm1/2)&=\beta_f^s \begin{pmatrix}
            \chi_\mp^\dagger(\theta_s) & \mp\alpha_f^s\chi_\mp^\dagger(\theta_s)
        \end{pmatrix}\equiv\bar u_{\pm} \ ,
    \end{align}    
\end{subequations}
with 
\begin{align}
    \chi_+(\theta)=\begin{pmatrix}
        \cos\frac{\theta}{2} \\ \sin\frac{\theta}{2}
    \end{pmatrix} \ , & &
    \chi_-(\theta)=\begin{pmatrix}
        -\sin\frac{\theta}{2} \\  \cos\frac{\theta}{2} 
    \end{pmatrix} \ ,
\end{align}
and $\beta_{i,f}^s=\sqrt{E_{i,f}^s+m_N}$, and $\alpha_{i,f}^s=|\mathbf{p}_{i,f}|/(E^s_{i,f}+m_N)$.

The Dirac structure of the electric term in the $s$-channel is given by
\begin{subequations}
\begin{align}
    \bar{u}_+\,\gamma_5\, u_+&= (\alpha_i^s-\alpha_f^s)\beta_i^s\beta_f^s\cos\frac{\theta_s}{2} \approx 0\ , \\
    \bar{u}_-\,\gamma_5\, u_-&=-\bar{u}_+\,\gamma_5\, u_+ \ ,\\
    \bar{u}_-\,\gamma_5\, u_+&= (\alpha_i^s+\alpha_f^s)\beta_i^s\beta_f^s\sin\frac{\theta_s}{2} \approx \sqrt{-t} \ , \\
    \bar{u}_+\,\gamma_5\, u_-&= \bar{u}_-\,\gamma_5\, u_+ \ .
\end{align}
\end{subequations}\label{eq:dirac-struc-s}

To summarize, in the limit $s\to \infty$, the non-vanishing $s$-channel helicity amplitudes are those for $\mu_i=\pm 1/2$ and $\mu_f=\mp 1/2$
\begin{equation} \label{eq:slimit-schannel}
A^\textrm{e}_{\mu_\gamma \pm \mp}(s,t)\approx -
   \left(2 e_\pi {\mu_\gamma} \sqrt{-t} \right)\frac{1}{ m_\pi^2-t}  \left( g_{\pi NN}  \sqrt{-t}\right) \ .
\end{equation} 
As expected, this corresponds to a factorized amplitude in the $t$-channel, \ie it can be written as the product of two vertices and the $t$-channel pion propagator. One vertex depends on the quantum numbers associated with the $\gamma\pi$, and the other corresponds to the $N\bar N$ vertex. The factor $\mu_\gamma$ is related to the parity symmetry of the $\gamma\pi$ vertex and each vertex contributes a factor $\sqrt{-t}$. 

Regarding the magnetic term in Eq.~(\ref{eq:amp-magnetic}), the nonzero structures in the $s$-channel frame are
\begin{subequations}
\begin{align}
    &\bar{u}_\pm\,\gamma_5\slashed{k}\slashed{\epsilon}({k,\pm 1})\,u_\pm \\ 
    &\qquad =-\sqrt{2}(1+\alpha_i^s)(1-\alpha_f^s)\beta_i^s\beta_f^s k_s\sin\frac{\theta_s}{2}\approx 0 \ , \nonumber\\
    &\pm\bar{u}_\mp\, \gamma_5\slashed{k}\slashed{\epsilon}({k,\pm 1})\,u_\pm \\ 
    &\qquad =\sqrt{2}(1+\alpha_i^s)(1+\alpha_f^s)\beta_i^s\beta_f^s k_s\cos\frac{\theta_s}{2}\approx\sqrt{2}s \ , \nonumber 
\end{align}   
\end{subequations}
and thus the two nonzero helicity amplitudes in the \mbox{$s\to\infty$} limit read
\begin{equation}\label{eq:slimit-schannel-mag}
    A^\textrm{m}_{\mu_\gamma=\pm,\mu_i=\pm,\mu_f=\mp}\approx\pm 2g_{\pi NN}(e_{N_i}-e_{N_f})
\end{equation}

\section{Born amplitudes in the $t$-channel}\label{app:Born-t}
In the $t$-channel frame, the $+\mathbf{\hat{z}}$ axis is chosen along the photon momentum and the nucleons in the $xz$ plane. The four-momenta are
\begin{subequations}
    \begin{align}
        k^\mu&=(k_t,0,0,k_t) \ , \\
        p_i^\mu&=(-E_N^t,-p_t\sin\theta_t,0,-p_t\cos\theta_t) \ , \\
        p_\pi^\mu&=(-E_\pi^t,0,0,k_t) \ , \label{eq:t-antinucleon}\\
        p_f^\mu&=(E_N^t,-p_t\sin\theta_t,0,-p_t\cos\theta_t) \ .
    \end{align}
\end{subequations}
We remark that the four-momentum of the pion, having negative energy, represents a pion in the initial state moving in the $-\mathbf{\hat{z}}$ direction. Similarly for the anti-nucleon. 
The photon has helicity $\lambda_\gamma=\pm 1$ and its polarization vector is 
\begin{equation}\label{eq:polvec_photon-t}
     \epsilon(k,\lambda_\gamma)=\frac{1}{\sqrt{2}}\left(0,- \lambda_\gamma,-i,0\right) \ ,
\end{equation}
and the nucleons have helicities \mbox{$\lambda_i=\pm 1/2$} and \mbox{$\lambda_f=\pm 1/2$}.

In Eq.~(\ref{eq:electric-3terms}) we have expressed the $s$-channel electric amplitude as a combination of three terms, each of which is independently gauge invariant and directly proportional to the electric charge of the exchanged particle. Upon evaluating this expression in the $t$-channel frame and using the definitions above, we obtain\footnote{The antinucleon four-momentum in \cref{eq:t-antinucleon} carries negative energy. Thus one can check that $u(p_i,\lambda)\propto v(-p_i, \lambda)$, where now the $v$ spinors are calculated with positive energy.}
\begin{align}\label{eq:Ael_tchannel}
    A^{\textrm{e}}_{\lambda_\gamma\lambda_i\lambda_f}&=-2g_{\pi NN}{\lambda_\gamma p_t\sin\theta_t}\frac{1}{s-u}  \\
     &\times \bigg[  2e_\pi+e_{N_i}\frac{t-m_\pi^2}{s-m_N^2} -e_{N_f}\frac{t-m_\pi^2}{u-m_N^2}\bigg]\nonumber \\ 
     &\times\,\bar{u}(p_f,\lambda_f)\,\gamma_5\,v(-p_i,\lambda_i) \ . \nonumber
\end{align}
In the $t$-channel, the photon energy $k_t$, the nucleon energies $E_N^t$ and momenta $p_t$, and the pion energy $E_\pi^t$ are
\begin{subequations}
    \begin{align}\label{eq:kt}
        k_t&=(t-m_\pi^2)/2\sqrt{t} \ , \\ 
        E_N^t&=\sqrt{t}/2 \ , \\
        p_t&=\sqrt{t/4-m_N^2} \ , \label{eq:pt} \\
        E_\pi^t&=(t+m_\pi^2)/2\sqrt{t} \ , 
    \end{align}
\end{subequations}
and the sine and cosine of the $t$-channel scattering angle, $\theta_t$, are
\begin{subequations}
    \begin{align}\label{eq:zt}
        \cos\theta_t&=\frac{s-u}{4k_tp_t} \ , \\
        \sin\theta_t&=\frac{\sqrt{\phi/t}}{2k_tp_t} \ .
    \end{align}
\end{subequations}
Based on Ref.~\cite{Mathieu:2015eia}, we adopt the Trueman and Wick convention~\cite{Trueman:1964zzb} in the $s$-channel physical region. In the large $s$ limit, $k_t$ and $p_t$ both become purely imaginary with a negative imaginary part, $\cos\theta_t\approx s/2k_tp_t$ is real and negative, and $\sin\theta_t\approx -is/2k_tp_t$ is a positive, purely imaginary quantity. 

From the expression of the electric amplitude at large $s$,
\begin{align}
    &A^\textrm{e}_{\lambda_\gamma\lambda_i\lambda_f}\approx 2g_{\pi NN}{\lambda_\gamma\sqrt{-t}}\bigg[e_\pi\frac{1}{t-m_\pi^2}  \\
    &\,+\frac12 e_{N_i}\frac{1}{s-m_N^2}+\frac12 e_{N_f}\frac{1}{s-m_N^2}\bigg]\bar{u}(p_f,\lambda_f)\,\gamma_5\,v(-p_i,\lambda_i)  \ , \nonumber
\end{align}
one finds the leading term to be precisely the piece of the nucleon exchanges that we added to the pion exchange to ensure its gauge invariance.

To evaluate the Dirac structure of the nucleons, given that the two nucleons are antiparallel in this frame (with the final-state nucleon along the direction $-\mathbf{\hat{p}}_t$), we use
\begin{subequations}\label{eq:spinors-t}
    \begin{align}
        v(-p_i,\pm 1/2)&=\beta_i^t\begin{pmatrix}
            -\alpha_i^t\chi_\mp(\theta_t) \\ \pm \chi_\mp(\theta_t)
        \end{pmatrix}\equiv v_\pm \ , \\
        \bar u(p_f,\pm 1/2)&=\beta_f^t \begin{pmatrix}
            \chi_\mp^\dagger(\theta_t) & \mp\alpha_f^t\chi_\mp^\dagger(\theta_t)
        \end{pmatrix} \equiv \bar u_\pm\ ,
    \end{align}    
\end{subequations}
with $\beta_{i}^t=\beta_{f}^t=\sqrt{E_N^t+m_N}$, and $\alpha_{i}^t=\alpha_{f}^t={p}_{t}/(E^t_{N}+m_N)$. The nonzero Dirac structures in the $t$-channel are
\begin{align}
    \bar{u}_+\,\gamma_5 \,v_+=-\bar{u}_-\,\gamma_5\, v_-=\sqrt{t} \ ,\label{eq:nonvanishingtchannelspinors}
\end{align}
and, in the limit $s\to\infty$, the nonzero $t$-channel helicity amplitudes are those with $\lambda_i=\lambda_f=\pm 1/2$
\begin{equation}   \label{eq:slimit-tchannel}
A^\textrm{e}_{\lambda_\gamma \pm\pm}(s,t)\approx i\,\left(2{\lambda_\gamma}e_\pi\sqrt{-t}\right)\frac{1}{m_\pi^2-t}\left(2\lambda_i g_{\pi NN}\sqrt{-t}\right)  \ .
\end{equation}

Crossing symmetry implies that the parity conserving combinations of the helicity amplitudes in the $s$- and $t$-channels are related by a rotation~\cite{Trueman:1964zzb,Collins:1977jy}. This is easy to see \eg in the large $s$ limit comparing the expressions in~\cref{eq:slimit-schannel,eq:slimit-tchannel}, 
\begin{equation}
    A^\textrm{e}_{\mu_\gamma +-}+A^\textrm{e}_{\mu_\gamma -+}=i\, (A^\textrm{e}_{\lambda_\gamma ++}-A^\textrm{e}_{\lambda_\gamma --})\delta_{\lambda_\gamma\mu_\gamma} \ .
\end{equation}
The phase is in agreement with~\cite{Mathieu:2015eia}.

The Dirac structures of the magnetic term in the $t$-channel, 
\begin{subequations}
    \begin{align}
        &\bar{u}_+\,\gamma_5\slashed{k}\slashed{\epsilon}(k,\lambda_\gamma)\,v_+=\bar{u}_-\,\gamma_5\slashed{k}\slashed{\epsilon}(k,-\lambda_\gamma)\,v_-  \\ 
        &\qquad =-2(E_N^t-\lambda_\gamma p_t)k_t\frac{\sin\theta_t}{\sqrt{2}} \approx -i\lambda_\gamma\frac{s}{\sqrt{2}}\ ,  \nonumber\\
        &\bar{u}_+\,\gamma_5\slashed{k}\slashed{\epsilon}(k,\lambda_\gamma)\,v_-=-\bar{u}_-\,\gamma_5\slashed{k}\slashed{\epsilon}(k,-\lambda_\gamma)v_+  \\ 
        &\qquad =-2Mk_t\frac{\cos\theta_t-\lambda_\gamma}{\sqrt{2}}\approx -i\frac{s}{\sqrt{2}} \ , \nonumber
    \end{align}
\end{subequations}
yield the following expressions for the helicity amplitudes in the $s\to\infty$ limit
\begin{subequations}
    \begin{align}
        A^\textrm{m}_{\lambda_\gamma\pm\pm}\approx\mp i\lambda_\gamma g_{\pi NN}(e_{N_i}-e_{N_f}) \ , \\
        A^\textrm{m}_{\lambda_\gamma\pm\mp}\approx\pm i g_{\pi NN}(e_{N_i}-e_{N_f}) \ ,
    \end{align}
\end{subequations}
and one finds that the crossing symmetry relations are also satisfied for the magnetic amplitudes
\begin{subequations}
    \begin{align}
        A^\textrm{m}_{\mu_\gamma +-}+A^\textrm{m}_{\mu_\gamma -+}&=i(A^\textrm{m}_{\lambda_\gamma ++}-A^\textrm{m}_{\lambda_\gamma --})\delta_{\lambda_\gamma\mu_\gamma} \ , \\
        A^\textrm{m}_{\mu_\gamma ++}-A^\textrm{m}_{\mu_\gamma --}&=-i(A^\textrm{m}_{\lambda_\gamma +-}+A^\textrm{m}_{\lambda_\gamma -+}) =0 \ .
    \end{align}
\end{subequations}

\section{Derivation of the partial wave of arbitrary spin $J$}
\label{app:exchange}
In this appendix, we give further details of the derivation of $t$-channel helicity amplitudes describing the exchange of arbitrary spin in the $t$-channel, and the spin summation. The kinematics in this frame are defined in \cref{app:Born-t}.

The polarization vector of the photon is given by \cref{eq:polvec_photon-t}.
For a massive spin-1 vector at rest,  the definition is
\begin{equation}\label{eq:polvec_massive-at-rest}
    \epsilon(\mathbf{0},\sigma)=\left\{\begin{aligned}\frac{1}{\sqrt{2}}&\left(0,\mp 1,-i,0\right) &\text{for }\sigma &= \pm 1\,,\\ &\left(0,0,0,1\right) & \text{for }\sigma &= 0\,.\\ \end{aligned}\right.
\end{equation}
For higher spin tensors, we use the relation
\begin{align}\label{eq:relation-eJ}
 \epsilon_{\nu_1\cdots \nu_{J}}(\helE) = \sum_{\sigma_1,\sigma_{J-1}} C^{J\helE}_{J-1\sigma_{J-1},1\sigma_1}
       \epsilon_{\nu_1\cdots \nu_{J-1}}(\sigma_{J-1})\epsilon_{ \nu_{J}}(\sigma_1) \nonumber
\end{align}
 with the Clebsch-Gordan coefficient defined as \mbox{$C^{J\helE}_{J-1\sigma_{J-1},1\sigma_1}\equiv\langle J,\helE| J-1,\sigma_{J-1}; 1,\sigma_1\rangle$}. For any generic four-vector $q$ in the $xz$ plane, the polarization vectors of Eqs.~(\ref{eq:polvec_photon-t}) and (\ref{eq:polvec_massive-at-rest}) satisfy 
\begin{equation}\label{eq:q-cdot-eps}
q^\mu\epsilon_\mu(\sigma)=-|\mathbf{q}|d^1_{\sigma 0}(\theta) \ ,
\end{equation}
where $d^1_{\sigma 0}(\theta)$ is the Wigner's (small) $d$-matrix element for $J=1$, and $\theta$ is the angle between $\mathbf{q}$ and the $+\mathbf{\hat{z}}$ direction. 
We follow the phase convention for the $d$-functions from Ref.~\cite{Varshalovich}.\footnote{The definition we adopt is given by \mbox{$d^J_{mm'}(\theta)=\left\langle jm\right| \exp(-iJ_y \theta)\left|jm'\right\rangle$\label{ft}}. This differs from the convention adopted in~\cite{Collins:1977jy,Mathematica}.} Some useful special values and symmetry relations are $d^J_{mm'}(0)=\delta_{mm'}$ and $d^J_{mm'}(\theta+\pi)=(-1)^{J-m'}d^J_{m-m'}(\theta)$.
For higher spin tensors, the following relation can be written
\begin{equation} \label{eq:pJeJ}
 q^{\nu_1} \cdots 
 q^{\nu_{J}} \epsilon_{\nu_1\nu_2\cdots\nu_{J}}
 (\helE) = (-|\mathbf{q}|)^J c_J d^J_{\helE 0}(\theta) \, 
\end{equation} 
with $c_J = J!\sqrt{2^J/(2J)!}$, which can be proved by induction. The coefficients $c_J$ satisfy the recurrence relation $c_J=C^{J0}_{J-10,10}c_{J-1}$, and \mbox{$C^{J\pm 1}_{J-1 0,1\pm 1}/C^{J0}_{J-10,10}=\sqrt{(J+1)/(2J)}$}. 
The sum of products of $d$-functions can be performed using
\begin{align}
    &\sum_{\sigma_{J-1},\sigma_1}C^{J\helE}_{J-1\sigma_{J-1},1\sigma_1}d^{J-1}_{\sigma_{J-1}0}(\theta)d^1_{\sigma_1\lambda}(\theta)= C^{J\lambda}_{J-10,1\lambda}d^J_{\helE\lambda}(\theta) \ .
\end{align}

With the above definitions and properties, we can compute the product of the vertices describing an unnatural exchange of spin $J$, \ie the combination of Eqs.~(\ref{eq:vertex-gammapi-unnat-traj}) and (\ref{eq:vertex-NNbar-unnat-traj-NF}), together with the Regge pole, $1/(J-\alpha_\pi(t))$. This gives the spin-$J$ amplitude,
\begin{widetext}
  \begin{align}\label{eq:amp-unnat-traj-NF}
    & A^J_{\lambda_\gamma\lambda_i\lambda_f}(s,t)=  \sum_{\helE}\frac{V^J_{\lambda_\gamma}(\helE)V^J_{\lambda_i\lambda_f}(\helE)}{J-\alpha_\pi(t)} \nonumber \\
&=-\frac{2\sqrt{2}e_{\pi} g}{J-\alpha_\pi(t)} (k \cdot p_{ \pi}) 
 \sum_{\helE}  P^{\nu'_1} \cdots P^{\nu'_{J}}\epsilon_{\nu'_1\cdots\nu'_J }(\helE) k^{\nu_1} \cdots k^{\nu_{J-1}}     
 \epsilon^*_{\nu_1\nu_2\cdots\nu_{J}}(\sigma_J) \epsilon^{\nu_J}(k,\lambda_\gamma)\,\bar u(p_f,\lambda_f)\,  \gamma_5 \,v(-p_{ i},\lambda_i) \nonumber
\\
      &=-\frac{2\sqrt{2}e_{\pi} g}{J-\alpha_\pi(t)} (k \cdot p_{ \pi}) 
 \sum_{\helE} P^{\nu'_1} \cdots P^{\nu'_{J}}\epsilon_{\nu'_1\cdots\nu'_J }(\helE)\sum_{\sigma_{J-1},\sigma_1}C^{J\helE}_{J-1\sigma_{J-1},1\sigma_1} k^{\nu_1} \cdots k^{\nu_{J-1}}     
 \epsilon^*_{\nu_1\nu_2\cdots\nu_{J-1}}(\sigma_{J-1})\big(\epsilon(k,\lambda_\gamma)\cdot\epsilon^*(\sigma_1)\big) \nonumber \\
  & \quad \times\,\bar u(p_f,\lambda_f)\,  \gamma_5 \,v(-p_{ i},\lambda_i) \,.
 \end{align}
We now use $P^{\nu'_1} \cdots P^{\nu'_{J}}\epsilon_{\nu'_1\cdots\nu'_J }(\helE)  = (-2p_t)^J c_J d^J_{\helE 0}(\theta_t+\pi)$ (as $\theta_t+\pi$ is the angle between $\mathbf{\hat{z}}$ and $-\mathbf{\hat{p}}_t$), and $k^{\nu_1} \cdots k^{\nu_{J-1}}\epsilon_{\nu_1\cdots\nu_{J-1} }(\sigma_{J-1})  = (-k_t)^{J-1} c_{J-1} d^{J-1}_{\sigma_{J-1} 0}(0)$ (as the photon is directed along the $\mathbf{\hat{z}}$ axis).
Finally, $\epsilon(k,\lambda_\gamma)\cdot\epsilon^*(\sigma_1) = -\delta_{\lambda_\gamma \sigma_1}$. We get
 \begin{align}
   & A^J_{\lambda_\gamma\lambda_i\lambda_f}(s,t)=   \\
 &=\frac{2\sqrt{2}e_{\pi} g}{J-\alpha_\pi(t)}(k\cdot p_{\pi})\sum_{\helE}\sum_{\sigma_{J-1}}C^{J\helE}_{J-1\sigma_{J-1},1\lambda_\gamma} (-2p_t)^Jc_Jd^J_{\helE 0}(\theta_t+\pi) (-k_t)^{J-1}c_{J-1}d^{J-1}_{\sigma_{J-1} 0}(0)\,\bar u(p_f,\lambda_f)\,  \gamma_5 \,v(-p_{ i},\lambda_i) \,. \nonumber
 \end{align}
 \end{widetext}
 We recall $d^J_{\helE 0}(\theta_t+\pi) = (-1)^{J} d^J_{\helE 0}(\theta_t)$ and $d^J_{\sigma_{J-1} 0}(0) = \delta_{\sigma_{J-1} 0}$.
Finally, using \cref{eq:nonvanishingtchannelspinors} and the recursion relations discussed above, we get
\begin{align} 
& A^J_{\lambda_\gamma\lambda_i\lambda_f}(s,t)=  \nonumber\\
 &=\frac{2\sqrt{2}e_{\pi} g}{J-\alpha_\pi(t)} \sqrt{t}(-2p_tk_t)^J\nonumber \\
 &\qquad\times\frac{C^{J\lambda_\gamma}_{J-1 0,1\lambda_\gamma}}{C^{J0}_{J-1 0,10}}(c_{J})^2d^J_{\lambda_\gamma 0}(\theta_t) \, 2\lambda_i\sqrt{t}\delta_{\lambda_i\lambda_f}
 \nonumber  \displaybreak \\ 
 &=\frac{2\sqrt{2}e_{\pi}g}{J-\alpha_\pi(t)}2\lambda_i\delta_{\lambda_i\lambda_f}c_J^2\sqrt{\frac{J+1}{2J}}(-2p_tk_t)^J t\, d^J_{\lambda_\gamma\lambda_i-\lambda_f}(\theta_t) \nonumber \\
 &= a^J_{\lambda_\gamma\lambda_i\lambda_f}(t) d^J_{\lambda_\gamma\lambda_i-\lambda_f}(\theta_t) \ .
\end{align}  

The Wigner $d$-function is written in terms of Jacobi polynomials as~\cite{Collins:1977jy}
  \begin{align}\label{eq:Wignerd_to_JacobiP}
    d^J_{\lambda\mu}(\theta)&=\xi_{\lambda\mu}\left[\frac{(J+M)!(J-M)!}{(J+N)!(J-N)!}\right]^{1/2} \\
    &\times \left(\frac{1-z}{2}\right)^{\frac{|\lambda-\mu|}{2}}\left(\frac{1+z}{2}\right)^{\frac{|\lambda+\mu|}{2}}P^{|\lambda-\mu||\lambda+\mu|}_{J-M}(z) \ , \nonumber 
\end{align}
with $M=\textrm{max}(|\lambda|,|\mu|)$, $N=\textrm{min}(|\lambda|,|\mu|)$ and $\xi_{\lambda\mu}=(-1)^{(\mu-\lambda-|\lambda-\mu|)/2}$, and $P_n^{ab}(z)$ the Jacobi polynomials,
\begin{align} \label{eq:hyp} 
P_{n}^{ab}(z) & = 
 \frac{\Gamma(a + n + 1 )}
{\Gamma(n+1)} \nonumber\\
&\times \mbox{}_2\tilde F_1\left(-n,1+a+b+n,a+1;\frac{1-z}{2}\right), 
\end{align} 
where $_2\tilde F_1$ is the regularized hypergeometric function.
We follow the phase convention for the $d$-functions from Ref.~\cite{Varshalovich}.\footref{ft}

\section{Details of the spin summation}
\label{app:sum}

The reduced amplitude $\mathcal{A}^{J\geq 2}_{j_p,j_z}$ in \cref{eq:Pregge} involves the following spin summation
\begin{align}
    \sum_{J=1,2,\ldots}\hspace{-2mm}&\frac{(2J+1)(J+1)}{J(J-\alpha)} \frac{j_p}{j_z}
\frac{J+j_z}{J+j_p} \nonumber \\
&\times (-\kappa)^J \frac{1}{2} \left[ 
P^{11}_{J-1}(z_t) - 
P^{11}_{J-1}(-z_t)\right] \ . 
\end{align}
By decomposing in partial fractions,
\begin{align} 
  &\frac{j_p}{j_z} \frac{(2J+1)(J+1)(J+j_z)}{J(J-\alpha)(J+j_p )} = 
  \frac{1}{J}\frac{1}{ (-\alpha) } \nonumber \\
  &\qquad + \frac{1}{J-\alpha} \frac{j_p(\alpha+j_z)(\alpha+1)(2\alpha+1)}{\alpha j_z(\alpha + j_p)}   \nonumber \\
  &\qquad +\frac{1}{J+j_p} \frac{(j_z - j_p)(1-j_p)(1-2j_p)}{j_z(j_p + \alpha) } 
  +\frac{2j_p}{j_z} \ ,
\end{align} 
we arrive to
\begin{align} \label{eq:fraction_expansion}
\sum_{J=1,2,\ldots} \hspace{-3mm}\cdots  &= \Bigg[\int_0^1   dy  \bigg(
 \frac{1}{ -\alpha }  + y^{-\alpha } 
 \frac{j_p(\alpha+j_z)(\alpha+1)(2\alpha+1)}{\alpha j_z(\alpha + j_p)} \nonumber \\
  &  +\, y^{j_p} \frac{(j_z - j_p)(1-j_p)(1-2j_p)}{j_z(j_p + 
 \alpha) }   \bigg) y^{J'} +\frac{2j_p}{j_z}\Bigg]  \nonumber \\
&\hspace{-12mm} \times \hspace{-2mm}  \sum_{J'=0,1,2,\ldots}  \hspace{-2mm}
(-\kappa)^{J'+1} \frac{1}{2} \left[ P^{11}_{J'}(z_t) - P^{11}_{J'}(-z_t)\right] .   
\end{align} 
where we have made use of the integral representation $(J+a)^{-1}=\int_0^1 dy\, y^{J-1+a}$ and relabelled the summation with $J'\equiv J-1$. The summation in \cref{eq:fraction_expansion} can be readily done using the generating function of the Jacobi polynomials,
\begin{align} \label{eq:sumJ}
&\sum_{n=0}^\infty  w^n \frac{1}{2} \left[ 
P^{11}_{n}(z) - P^{11}_{n}(-z)\right]  \\ 
& = 2\left[  \frac{1}{G(z,w)}\frac{1}{
 (1 + G(z,w))^2 - w^2   }  - (z \to -z) \right] \ , \nonumber
\end{align} 
with $G(z,w)  = \sqrt{1 + 2 w z 
 + w^2 }$, and gives the expression in \cref{eq:Pregge_eff}.

One can change the integration variable, $y \to x \equiv  \kappa  z_t y = (s-u) R^2  y/4$.
In the physical region of the $s$-channel ($s>0$ and $u<0$), the second term in Eq.~(\ref{eq:sumJ}) has a branch point singularity. Therefore, we evaluate the integral of this term in the unphysical region ($s<0$ and $u>0$) and analytically continue the result to the physical region.

In the high-energy limit, we have approximately $x\approx s R^2  y/2 \approx - uR^2 y/2$, and $G(x)\approx\sqrt{1+2x}$, which gives
\begin{align}\label{eq:int}
 \int_0^1 dy \, &\cdots \to -\frac{1}{z_t} \int_0^{sR^2/2}dx \,
\bigg[
 \frac{1}{ -\alpha } \\
& +\left(\frac{sR^2}{2}\right)^{\alpha} \frac{1}{x^{\alpha } }
 \frac{j_p(\alpha+j_z)(\alpha+1)(2\alpha+1)}{\alpha j_z(\alpha + j_p)}  \nonumber \\
 &+ \left(\frac{2}{sR^2}\right)^{j_p} x^{j_p} \frac{(j_z - j_p)(1-j_p)(1-2j_p)}{j_z(j_p + \alpha) }  \bigg] \nonumber \\
 & \times 2   \left(\frac{1}{
 \sqrt{1 + 2x}(1+\sqrt{1+2x})^2  } - (s \to u) \right) \ . \nonumber 
\end{align} 
The contribution from the Regge pole alone, as opposed to the fixed pole at $J=-j_p$, is obtained by taking the limit $j_z,j_p \to \infty$. We use 
\begin{align}
   &\int_0^{sR^2/2}dx\frac{(a+1)(2a+1)x^{-a}}{\sqrt{1+2x}(1+\sqrt{1+2x})^2} = \frac{2^{a}\sqrt{\pi}\Gamma(a+3/2)}{\Gamma(a)\sin(\pi a)} \nonumber \\
    &-\frac{(a+1)}{\sqrt{2}(sR^2/2)^{a+1/2}} {}_2F_1\left(-\frac12,\frac12+a;\frac32+a;-\frac{1}{sR^2}\right) \nonumber \\
    & 
   +\frac{(1+2a)}{2(sR^2/2)^{a+1}}
      -\frac{(a+1)(2a+1)}{2\sqrt{2}(3+2a)(sR^2/2)^{a+3/2}} \nonumber \\
    &  \times{}_2F_1\left(\frac12,\frac32+a;\frac52+a;-\frac{1}{sR^2}\right) .
\end{align}
For $-1/2<a<1$, the upper integration limit can be taken to infinity, 
\begin{align} 
\int_0^\infty dx\, &\frac{(1+a
)(1+2a)x^{-a} }{\sqrt{1+2x}
(1+\sqrt{1+2x})^2}= \frac{2^{a}\sqrt{\pi}
\Gamma(a+3/2)}{\Gamma(a)\sin\pi a} \nonumber \\
&= 2^{a}  \Gamma(1 - a) \frac{\Gamma(a+3/2)}{\sqrt{\pi}} \ .
\end{align}
This yields
\begin{align}\label{eq:integral_limit}
& \int_0^1 dy \, \cdots \to  
   -\frac{1}{z_t} 
 \frac{1}{-\alpha}\Bigg[\left(1  - (sR^2)^\alpha 
\frac{2\sqrt{\pi} \Gamma(\alpha+3/2)  }{\Gamma(\alpha)\sin\pi\alpha} \right) \nonumber \\
& \qquad+\left(1  - (uR^2)^\alpha 
\frac{2\sqrt{\pi} \Gamma(\alpha+3/2)  }{\Gamma(\alpha)\sin\pi\alpha} \right)\Bigg] \nonumber  \displaybreak \\ 
&=-\frac{1}{z_t} 
 \frac{1}{-\alpha}\
 \Bigg[\left(1  - (sR^2)^\alpha 
\frac{2\sqrt{\pi} \Gamma(\alpha+3/2)  }{\Gamma(\alpha)\sin\pi\alpha} \right) \nonumber \\
& \qquad+  \left(1  -e^{-i\pi\alpha} (sR^2)^\alpha 
\frac{2\sqrt{\pi} \Gamma(\alpha+3/2)  }{\Gamma(\alpha)\sin\pi\alpha} \right)\Bigg] \nonumber   \\ 
&=-\frac{2}{z_t}\frac{1}{-\alpha} 
\left(1  - (sR^2)^\alpha 
\frac{\tau 2\sqrt{\pi}}{\sin\pi\alpha}\frac{ \Gamma(\alpha+3/2)  }{\Gamma(\alpha)} \right) \ ,
\end{align}
where $\tau = \left[ 1 + \exp(-i\pi\alpha) \right] /2$ is the signature factor and allows us to write the reduced amplitude in \cref{eq:Pregge_Jlarge}.

\bibliographystyle{apsrev4-1}
\bibliography{refs}
\end{document}